\documentclass[12pt]{article}
\pdfoutput=1
\usepackage{amsmath}
\usepackage{enumerate}
\usepackage{enumitem}
\usepackage{amsfonts}
\usepackage{graphicx}
\usepackage{bm}
\usepackage{bbm}
\usepackage{natbib}
\usepackage[T1]{fontenc}

\usepackage{graphics}
\usepackage{float}
\usepackage{times}
\usepackage{hhline}
\usepackage{caption}
\usepackage{subcaption}
\usepackage{xcolor}
\usepackage{pdflscape}
\usepackage[margin=1in]{geometry}
\usepackage{setspace}
\newcommand{\m}[1]{\mathbf{\bm{#1}}}

\newcommand{\bfone}{{\bf 1}}

\newcommand{\bfphi}{{\bm \phi}}

\newcommand{\bfdelta}{{\bm \delta}}
\newcommand{\bfbeta}{{\bm \beta}}

\newcommand{\bfSigma}{{\bm \Sigma}}

\def\bfx{\mathbf{x}}

\def\bfP{\mathbf{P}}
\def\bfX{\mathbf{X}}

\def\bfA{\mathbf{A}}

\def\bfQ{\mathbf{Q}}

\def\bfM{\mathbf{M}}
\def\bfI{\mathbf{I}}

\def\inv{^{-1}}

\def\inv{^{-1}}

\usepackage[affil-it]{authblk}
\makeatletter
\def\@maketitle{%
  \newpage
  \null
  \vskip 2em%
  \begin{center}%
  \let \footnote \thanks
    {\Large\bfseries \@title \par}%
    \vskip 1em%
    {\normalsize
      \lineskip .2em%
      \begin{tabular}[t]{c}%
        \@author
      \end{tabular}\par}%

    {\small \@date}%
  \end{center}%
 \vskip 1em
  \par
}

\title{Modeling Efficiency of Foreign Aid Allocation in Malawi}
\author[1]{Philip A.\ White\thanks{philip.a.white@duke.edu}}\affil[1]{Department of Statistics, Duke University, Durham, NC, USA}
\author[2]{Candace Berrett\thanks{cberrett@stat.byu.edu}}\affil[2]{Department of Statistics, Brigham Young University, Provo, UT, USA}
\author[2]{E.\ Shannon Neeley-Tass\thanks{tass@stat.byu.edu}}
\author[3]{Michael G. Findley\thanks{mikefindley@utexas.edu}}\affil[3]{Department of Government, University of Texas at Austin, Austin, TX, USA}

\begin{document}  
\maketitle
\abstract{
The Open Aid Malawi initiative has collected an unprecedented database that identifies as much location-specific information as possible for each of over 2500 individual foreign aid donations to Malawi since 2003.  
Ensuring efficient use and distribution of that aid is important to donors and to Malawi citizens.  However, because of individual donor goals and difficulty in tracking donor coordination, determining presence or absence of efficient aid allocation is difficult.  We compare several Bayesian spatial generalized linear mixed models to relate aid allocation to various economic indicators within seven donation sectors.  We find that the spatial gamma regression model best predicts current aid allocation. Using this model, first we use inferences on coefficients to examine whether or not there is evidence of efficient aid allocation within each sector.  Second, we use this model to determine a more efficient aid allocation scenario and compare this scenario to the current allocation to provide insight for future aid donations.  
}


\section{Introduction} \label{sec:intro}

Each year, wealthy countries and international organizations send billions of dollars in foreign aid  to developing countries across the world. The Organization for Economic Co-operation and Development (OECD) estimates that in 2015 the United States alone donated 31.1 {billion} dollars in development assistance and that other developed countries are donating anywhere between 700 million and 19 billion US dollars \citep[][]{oecd2015}. 
	Aid giving is thus big business, and we need to understand better whether the business operates in the service of its so-called clients in recipient countries or rather in service of its own interests.

Donors' motives are varied, ranging from pursuing strategic interest \citep{AlesinaDollar:2000} to trying to facilitate positive social and economic outcomes. 	And yet, with so much spent on development assistance, donors, governments, and citizens may want these donations to be allocated efficiently --- providing money to places and situations with the most need. Even where strategic interest concerns are high, and therefore aid may not be allocated with need in mind, foreign aid can affect peoples' lives in non-trivial ways, for better or for worse. Thus, stakeholders in both recipient and donor countries deserve to know whether aid is in fact allocated in ways that respond to and meet the needs of recipient countries.  For example, donations meant for improving education should ideally be sent to those places within the country where education is suffering the most.  

Several initiatives have been introduced to improve aid targeting and coordination, including the Paris Declaration of 2005 which called for better alignment between government objectives and donor aid allocation \citep{aced2009, FindleyEtAl:2016}.  However, communication among donors and recipients, as well as donors and other donors, in order to track and efficiently allocate the vast array of projects has proven nearly impossible. Moreover, there has been a dearth of scholarly attempts to track the complexity of aid allocation and so both academics and policy makers have a poor understanding of the patterns of foreign aid allocation. Because so much hangs in the balance, answers to the question of efficient aid allocation are sorely needed. 

In 2011, the government of Malawi, together with the University of Texas at Austin's Climate Change and African Political Stability team,  began efforts to collect and display donor information at the subnational level \citep{FindleyEtAl:2011,weav2014}. This initiative --- Open Aid Malawi --- resulted in an unprecedented and unique database containing as much location-specific information as possible for \emph{each} individual donation to Malawi since 2003 \citep{pera:etal:2012}.  This database contains over 2500 tracked donations all at the subnational level, to date the largest of its kind. Such a large and detailed resource makes it possible to examine donor aid across various project sectors.  

Malawi is a particularly interesting country for examining aid allocation.  
Malawi's economy, education, and public health, are some of the least developed anywhere in the world. Specifically, they have an infant mortality rate of 69.3 for every 1,000 births, 50.7\% of the population live below the poverty line,  a 68.2\% literacy rate, and agricultural goods are their leading export \citep{NSO2015}.
In an effort to improve circumstances in Malawi, a large number of foreign donors support economic, health, and educational development and improvement plans in Malawi. In contrast to most highly impoverished countries, Malawi has experienced almost no violence, making it an ideal location to study aid allocation patterns as the presence of violence makes stable aid allocation patterns difficult. Put differently, given its political stability Malawi is a most likely case to observe donors responding to need, and we thus explore in this paper the extent to which donors in fact allocate based on indicators of need.

Because so much aid is sent to Malawi and because of its need for hastened development, the question of whether aid is allocated efficiently is paramount both for taxpayers in donor countries as well as citizens affected by aid in recipient countries. We are specifically interested in quantifying how assistance is being allocated within the country to determine whether sectoral aid matches sectoral need. In particular, we focus on seven project sectors or donation types:  (1) Agriculture; (2) Education; (3) Governance; (4) Health; (5) Rural Development [RD]; (6) Roads, Public Works, and Transportation [RPT]; and (7) Water, Sanitation, and Irrigation [WSI].

Using district-specific economic, social, and demographic data (hereafter referred to as ``economic indicators''), we model foreign aid allocation in all districts for these project sectors. 
While we do not expect the economic indicators to be the sole driving forces behind aid allocation, by donors own admission they should be key factors in determining whether or not aid is being allocated efficiently.  Therefore, our goals are, first, identifying high-level economic indicators that are related to aid allocation while accounting for and making use of spatial dependence; second, determining the ability of our model to capture a true relationship between aid distribution and the economic indicators; and third, to propose an efficient allocation scenario and its discrepancy from the actual allocation in order to inform future donations.  

Analyses attempting to measure donor coordination in Malawi using this novel data set have begun.  For example, comparing donations before and after the Paris Declaration of 2005 with various economic and demographic indicators, \cite{Nunn2015} found no evidence of increased donor coordination across sectors.  \cite{De2015} used instrumental variable techniques to show evidence of improved economic conditions at the local level from aid donations, emphasizing the need and ability for donors to coordinate their efforts using local economic indicators.  Using a model to accurately capture the skewness present in aid data, we seek to significantly expand on these analyses by exploring the extent of the spatial dependence of the different sectors and to account for this spatial dependence to make inferences about relationships between donations and economic indicators at the local level. 

Common analysis tools in the literature that examine aid allocation on this and other data sets include multiple regression \citep{feeny2007, Nunn2015}, probit and Tobit regression \citep{collier2002, dietrich2013, De2015}, poisson regression \citep{masaki2017}, structural equation models and two-stage least squares \citep{collier2002, dietrich2015}, and mixed effects models \citep{hodler2014, jablonski2014, briggs2017}.
As geocoded aid data becomes more readily available, models that account for spatial dependence are becoming more common.  For example, \cite{vanweezel2015} used a Bayesian normally-distributed mixed model with a spatial random effect to examine the relationship between foreign aid and civil conflict.  \cite{nunnen2017} study the motivation (i.e. need, merit, or politics)
 for aid allocation in India using a spatial two-stage least squares.   \cite{runfola2017} developed a
 methodology which they call geoSIMEX to account for different levels of spatial imprecision that arise when aid allocation data is reported at 
various spatial resolutions.  Our model helps build on and expand the growing literature that accounts for
spatial dependence in the modeling of foreign aid allocation data.


In this manuscript, we examine donation allocation within Malawi at the subnational level using a Bayesian spatial gamma regression model.  We describe the data in more detail in Section \ref{sec:data}.  In Section \ref{sec:model} we explicitly define the model used for analysis as well as the comparison models.  In Section \ref{sec:params} we provide an in-depth analysis of the various model parameters and their interpretations and implications for aid allocation in Malawi, contrasting these results with a more ideal allocation scenario in Section \ref{sec:suggestions}.  Finally, we provide concluding remarks in Section \ref{sec:conclusion}.

\section{Data} \label{sec:data}

\begin{figure}[t]  
\vspace*{-.5in}
\begin{center}
  \begin{subfigure}[b]{.32\textwidth}
  \centering
    \includegraphics[width=1\textwidth]{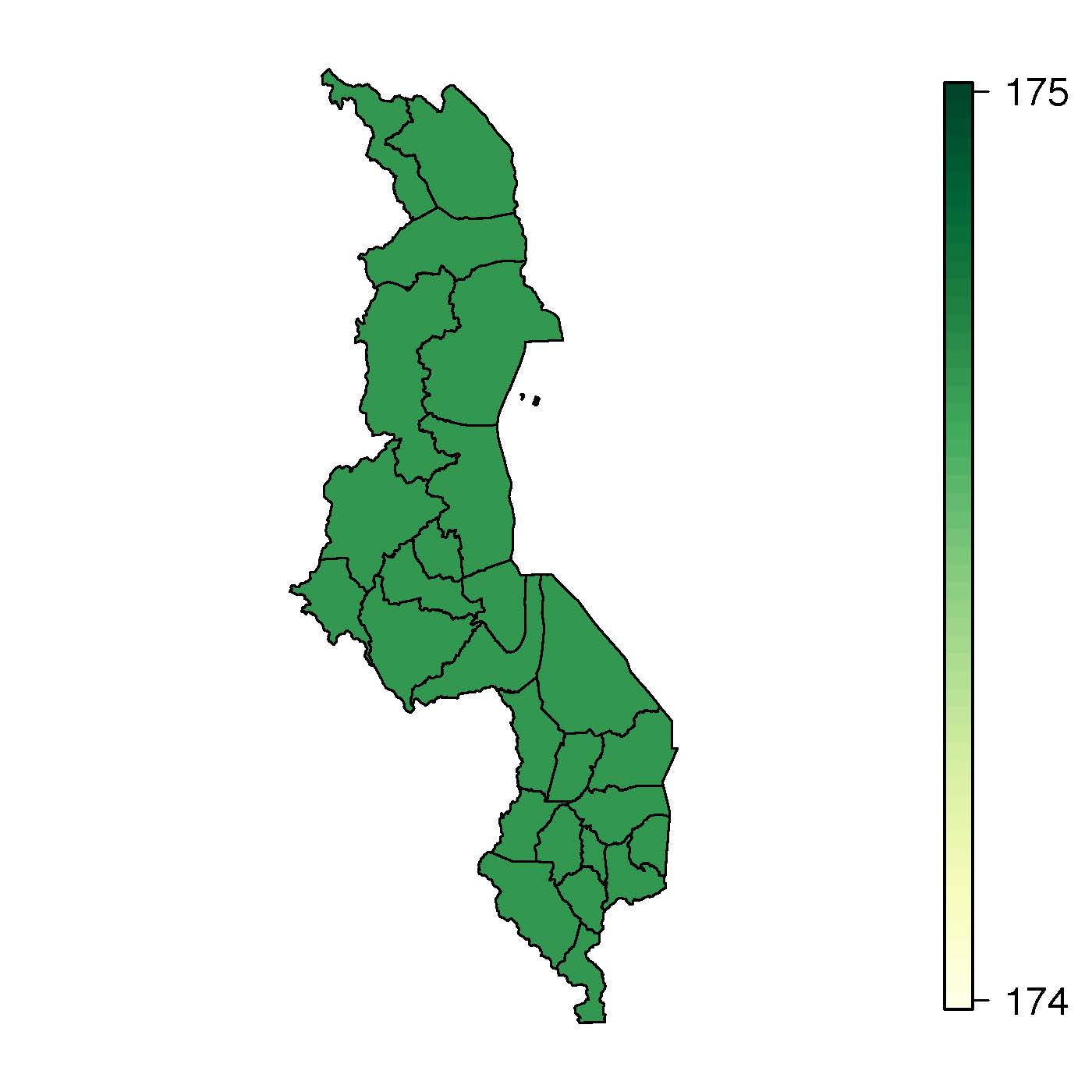} 
        \subcaption{Country Level Donations}\label{fig:country}
    \end{subfigure}
      \begin{subfigure}[b]{.32\textwidth}
  \centering
    \includegraphics[width=1\textwidth]{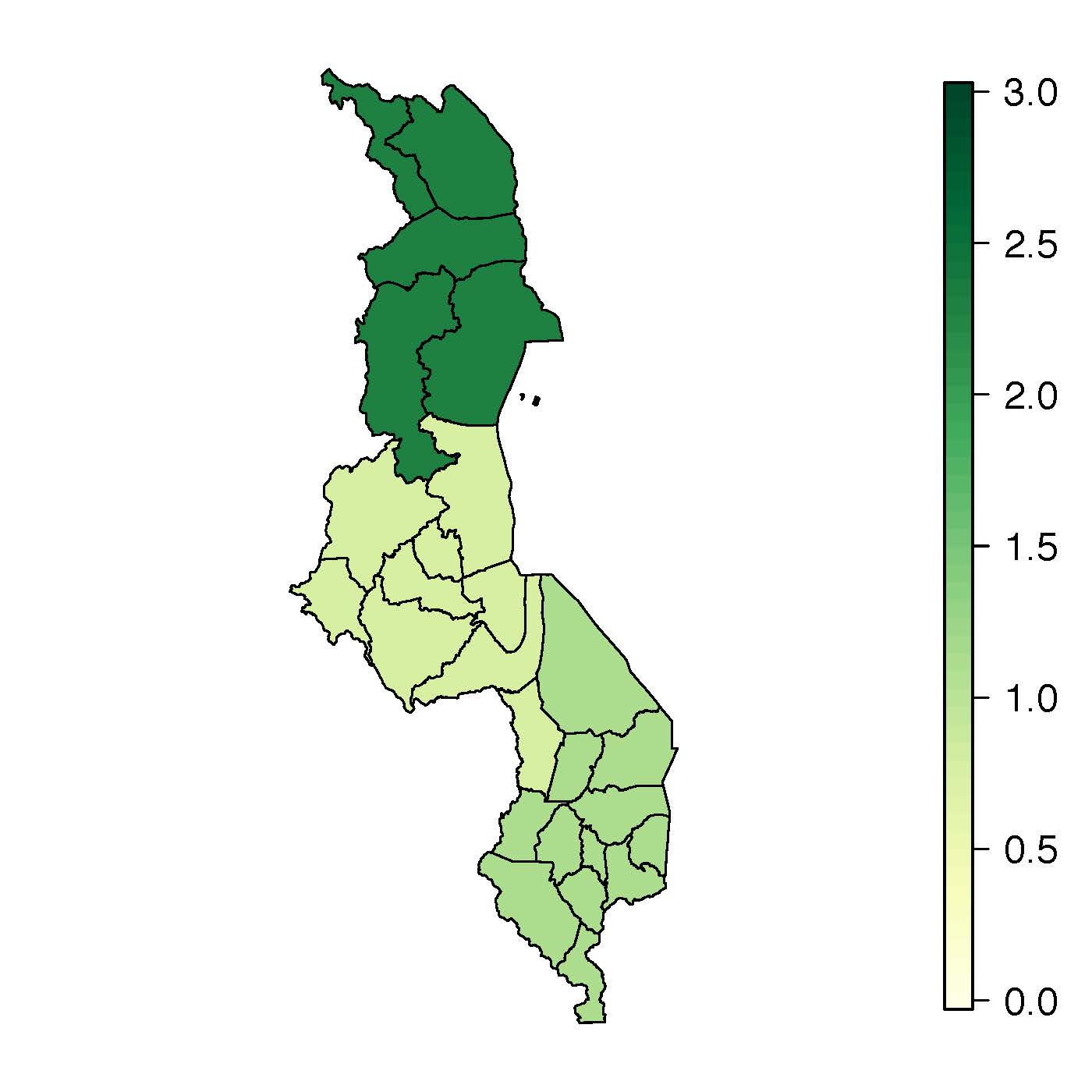}
    \subcaption{Region Level Donations} \label{fig:region}
    \end{subfigure}      
      \begin{subfigure}[b]{.32\textwidth}
        \centering
    \includegraphics[width=1\textwidth]{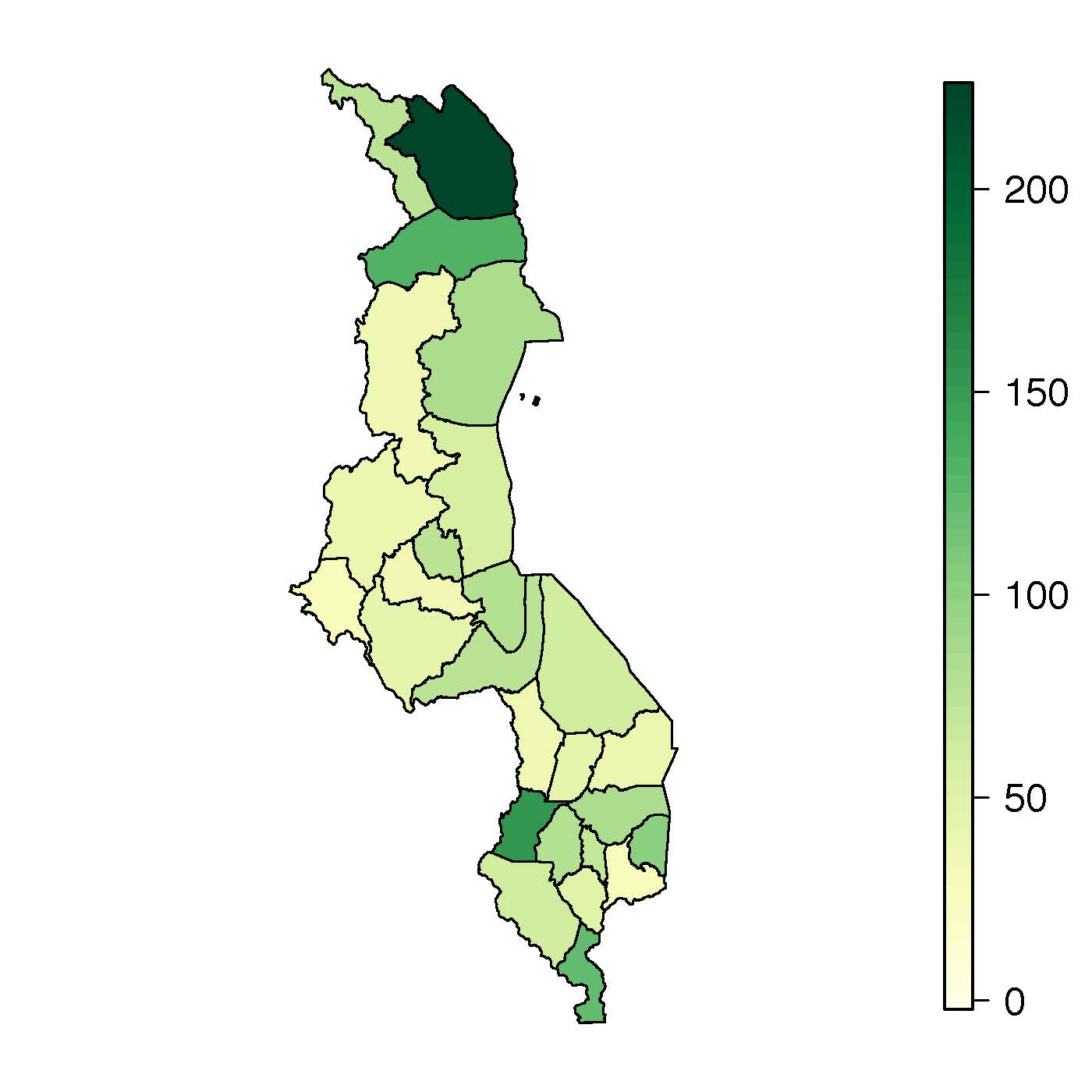} 
        \subcaption{District Level Donations}\label{fig:district}
    \end{subfigure}
      \begin{subfigure}[b]{.32\textwidth}
  \centering
    \includegraphics[width=1\textwidth]{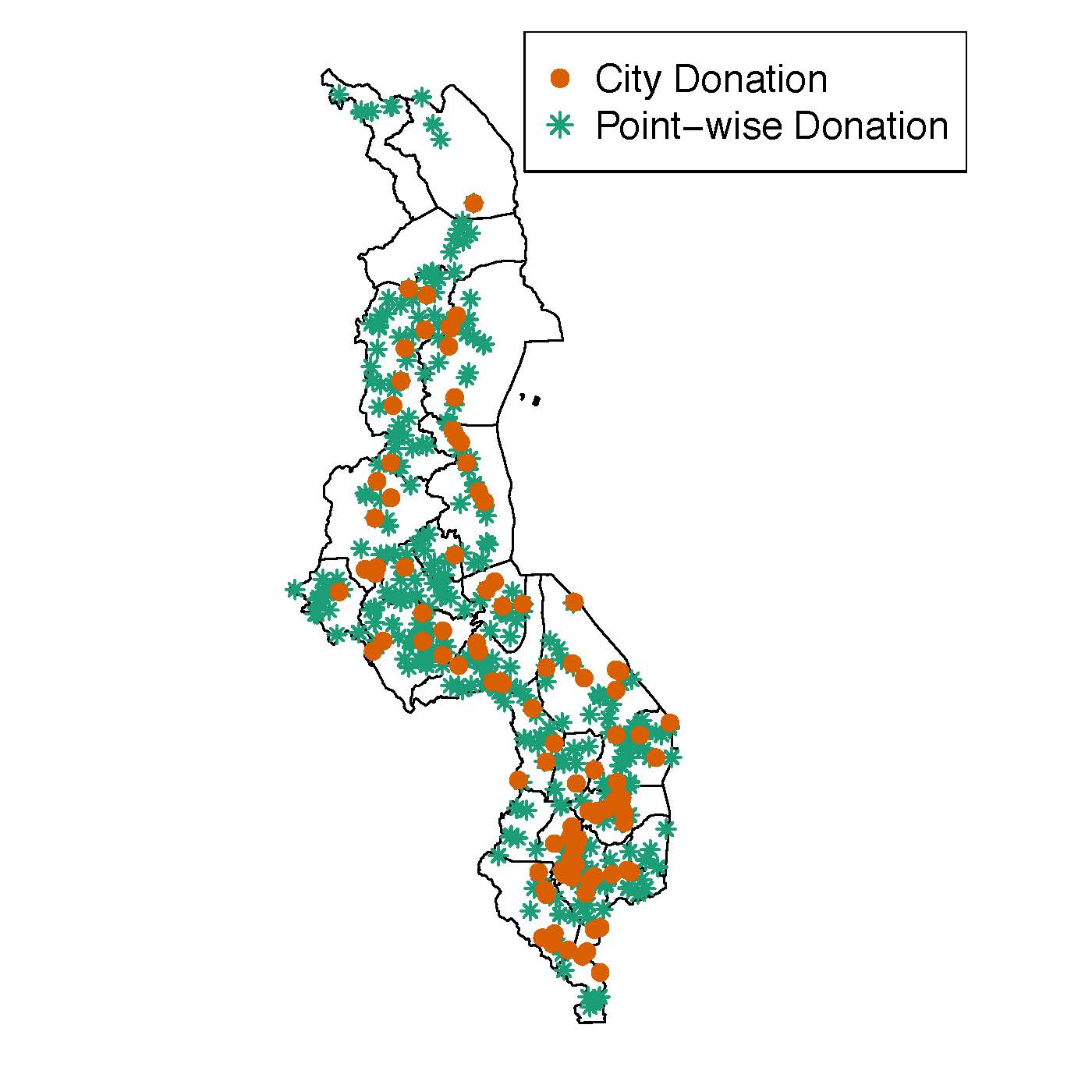}
    \subcaption{Point Level Donations} \label{fig:point}
    \end{subfigure}      
      \begin{subfigure}[b]{.32\textwidth}
  \centering  
    \includegraphics[width=1\textwidth]{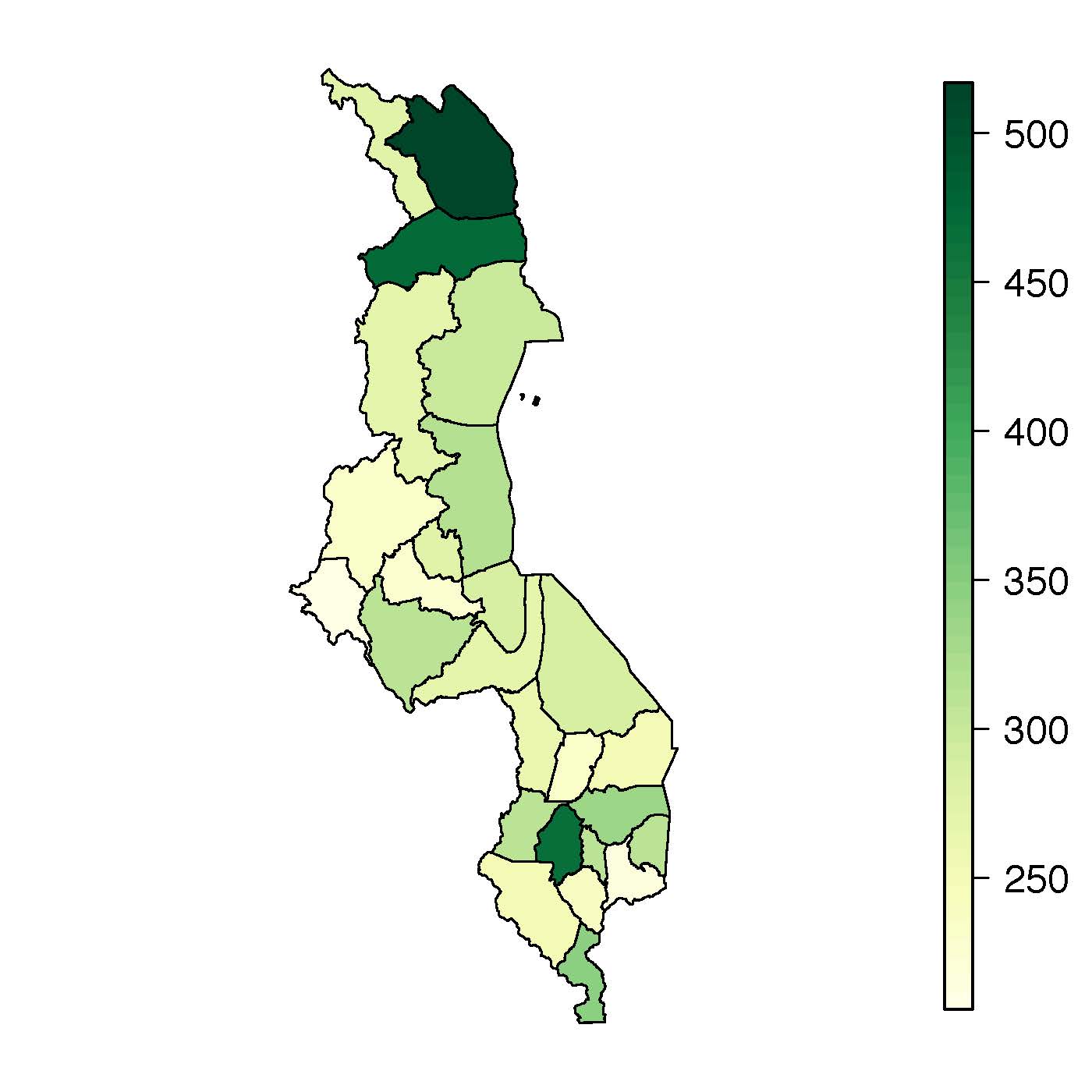}
    \subcaption{Summed Donations} \label{fig:sum}
    \end{subfigure}     
          \begin{subfigure}[b]{.32\textwidth}
    \centering
    \includegraphics[width=1\textwidth]{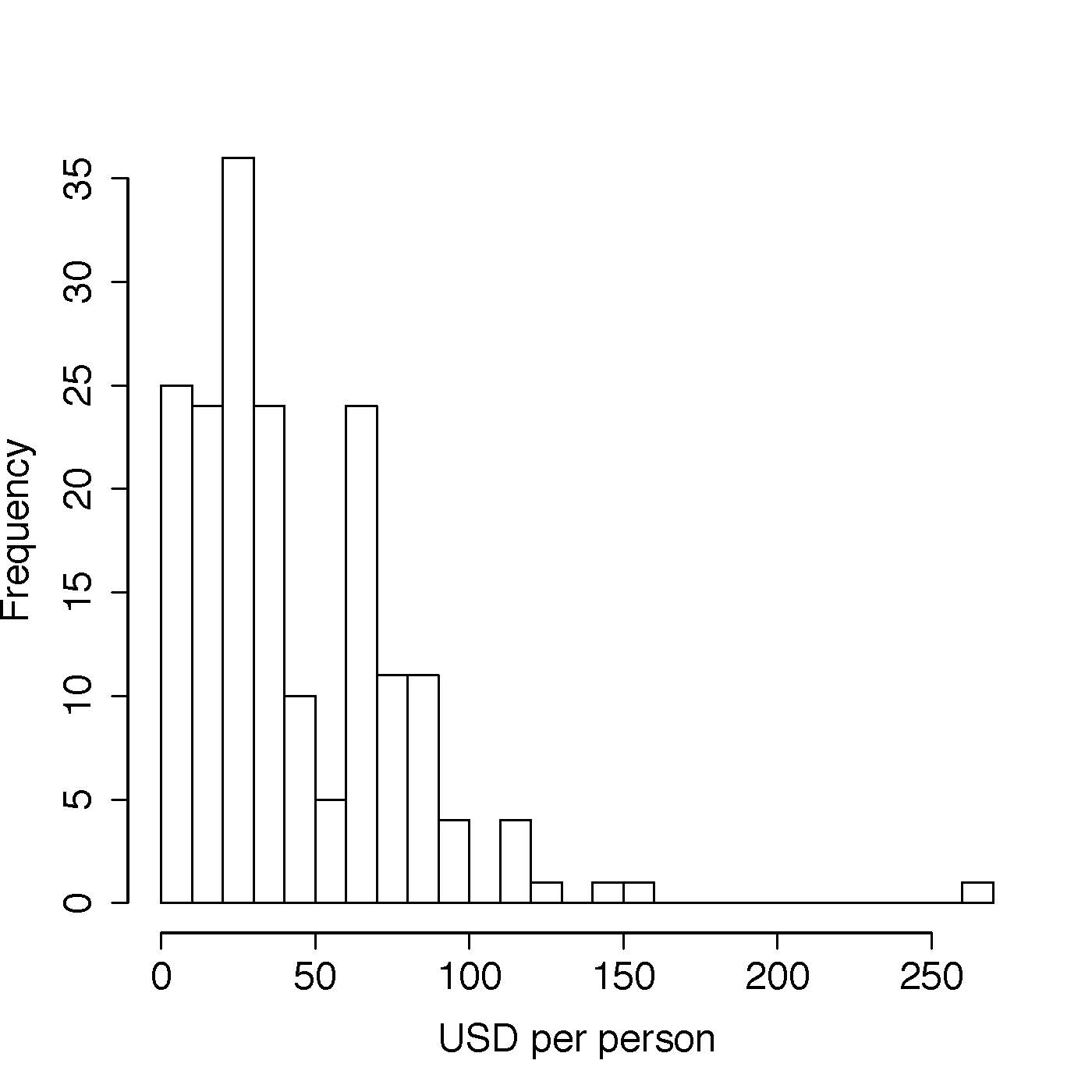}
    \subcaption{Histogram of Donations} \label{fig:hist}
    \end{subfigure}     
   \caption{Money per person for donations at the various levels of available location information.  Figures (a) -- (c) show maps of money per person for (a) donations made to the entire country, (b) donations made to the individual northern, central, and southern regions, and (c) donations made to the individual districts.  Figure (d) shows a map of the locations of city and point-wise donations.  Figure (e) shows the total money per person within each district.  Finally, (f) shows a histogram of the donations per person within each project sector and district.}\label{fig:data}
  \end{center}
\end{figure}

The Open Aid Malawi database \citep{pera:etal:2012} contains over 2500 donations spanning the years 2003 to 2016.  Each donation total (in US dollars; USD) corresponds to a specific project and its intended project sector (Agriculture; Education; Governance; Health; Rural Development; Roads, Public Works, and Transport; and Water, Sanitation, and Irrigation).  The database also contains the spatial location and ``precision'' or level at which a project's donation was designated (e.g., country, region, district, city). Figures \ref{fig:country}-\ref{fig:point} illustrate the different precisions at which a project's donation could be assigned.  For example, Figure \ref{fig:region} shows the total money per person assigned to the three regions (North, Central, South) in Malawi, while Figure \ref{fig:point} shows the locations of project donations assigned to a city (orange circle points) or a more specific latitude-longitude (green stars).

In addition to the donation data, we also have population and  economic indicator data for each district from the 2008 Malawi census \citep{NSO2008}.  These economic indicators include: proportion of people in poverty, average distance of people to schools (km), number of recently (within two weeks) injured people, food spending as a percentage of all expenses, mean area cultivated per household, and proportion of population with electric lighting.  Descriptions of the variables are in Table \ref{tab:covariates}; all covariates were standardized across districts to have a mean zero and variance one.  However, we do not have these values for the three smallest districts -- Likoma (a small island in Lake Malawi), Neno (split in 2003 from Mwanza), and Area under National Administration (various locations across the country).  To accommodate this lack of information, we combined the donations to Neno and Mwanza (as they only recently split) and removed the other two districts from the analysis.

Because our economic indicator data is only available at the district level, we aggregated all the donation data to the district level.  For the country- and region-level donations, we assumed these were dispersed proportionally among the districts according to district population.  Additionally, each individual project could have several subprojects and each subproject could be assigned to a different location.  However, only the total project donation amount is reported in the database and not the donation for the specific subproject.  In this case, we divided the total project donation amount equally among the subprojects.  Our final donation data is the total donation sum for each district in USD by project sector. Figure \ref{fig:sum} shows the total donations per person summed across project sectors.  We expect that more money should be sent to those districts with more people, thus we use dollars per person as the response variable by dividing the district donation total for each project sector by the district population.  

For the dataset used in the analysis, the total sample size is $N = 182$ since the data contain dollars donated per person in 26 districts for seven project sectors. Figure \ref{fig:hist} shows a histogram of all 182 observed dollars per person for each sector and district, where the most extreme value is \$265.42 per person.  

\begin{landscape}
\begin{table}
\small
\caption{Names and descriptions of the data variables used in the analysis.}\label{tab:covariates}
\begin{tabular}{clll}
\hline
&Variable & Abbreviation & Description \\\hline\hline
\multicolumn{2}{l}{\emph{Sectors}} &\\
& Agriculture & Agriculture & Donations assigned to development of agriculture\\
& Education & Education & Donations assigned to education development\\
& Governance & Governance & Donations assigned to both economic and democratic governance\\
& Health & Health & Donations assigned to for development of health initiatives\\
& Integrated Rural Development & RD  & Donations assigned for rural development\\
& Roads, Public Works, and Transportation & RPT & Donations assigned to developing roads, transportation, and other public works \\
& Water, Sanitation, and Irrigation & WSI & Donations assigned to water, sanitation, and irrigation development\\
\hline
\multicolumn{2}{l}{\emph{Economic Indicators} } & \\
& Proportion in Poverty & Poverty & Percent of population living below the national poverty line\\
& Distance to School &  Dist to School & Average distance (km) across households to the nearest school in each district  \\
& All Injured & All Injured & Number injured or ill in the previous 2 weeks (as of the 2008 census) in each district\\
& Food as \% of Expenditures & Food \% & Average percent of household expenditures spent on food in each district\\
&  Mean Land Cultivated per House & Land Cultivated &  Average area cultivated per household (in hectres) in each district \\
& Electric Lighting Proportion & Electric \% & Percent of households in the district with electric lighting \\
& Population & Population & Population of each district \end{tabular}
\end{table}
\end{landscape}

\section{Model}\label{sec:model}

\subsection{Model Definition}\label{sec:modeldef}

We expect some quantities, like agricultural output, public health outcomes, or non-urban areas, to be grouped spatially; therefore, it is plausible that accounting for spatial dependence could improve model performance in terms of prediction and inferences. The normal linear regression model with spatial random effects is a standard model for spatially-distributed quantities \citep{banerjee2014}. Because our data is strictly positive and right-skewed, we consider more general models.  Following  \cite{diggle1998}, we make use of a spatial generalized linear mixed model (SGLMM).   

Let $Y_{ij}$ be the dollars per person (USD) for the $i^{th}$ district ($i = 1, \dots, n$, where $n=26$) and the $j^{th}$ project sector ($j = 1, \dots, J$, where $J = 7$).  We model
\begin{align*}
Y_{ij} | \mu_{ij}, \theta \sim f(y_{ij} | \mu_{ij}, \theta),
\end{align*}
where $\mu_{ij} = E(Y_{ij} | \bfbeta_j, \phi_{ij}, \theta)$ and $\theta$ is an applicable scale parameter (for example, for the normal distribution, this is the common variance parameter).  We model the mean using a link function, $g(\cdot)$, 
\begin{align}
g(\mu_{ij}) = \bfx_{ij}' \bfbeta_j + \phi_{ij}, \label{eq:lm}
\end{align}
where $\bfx_{ij}$ is a $k \times 1$ vector of covariates (economic indicators) for the $i^{th}$ district and the $j^{th}$ project sector, $\bfbeta_j$ is a $k \times 1$ vector of coefficients for each sector, and $\phi_{ij}$ is a spatial random effect where each sector has its own spatial dependence structure.  Because we expect a combination of the economic indicators to be related to aid allocation within each sector, we fit a sector-specific coefficient for each of the seven economic indicators (including population) as well as a sector-specific intercept to account for the different amounts of aid appointed to each sector.  Therefore, there are $k=8$ coefficients for each sector, or 56 total coefficients in our model.  

Following \cite{hugh:hara:2013}, we account for potential spatial confounding by using the Moran basis on the spatial random effect.   
 Let $\bfphi_j = (\phi_{1j}, \dots, \phi_{26,j})'$ be the vector of spatial random effects for sector $j$ and model this with an intrinsic conditionally autoregressive \cite[CAR; see, e.g.,][]{besag1974, banerjee2014} prior:
 \[
 \pi(\bfphi_j | \sigma^2) \propto \left(\frac{1}{\sigma^2}\right)^{\mbox{rank}(\bfQ)/2} \exp\left\{-\frac{1}{2\sigma^2}\bfphi_j'\bfQ\bfphi_j\right\},
 \]
where $\sigma^2$ is a variance parameter and $\bfQ = \mbox{diag}(\bfA\bfone) - \bfA$ is a precision matrix where $\bfA$ is the $n \times n$ binary neighborhood matrix such that $A_{ik} = 1$ for $i \ne k$ if districts $i$ and $k$ share a border and $A_{ik} = 0$ otherwise, and $\bfone$ is an $n \times 1$ vector of ones. As detailed by \cite{reic:etal:2006}, the spatial random effect may be correlated with spatially-dependent covariates, creating collinearity between the model coefficients, $\bfbeta$, and the spatial random effect, $\bfphi$, inhibiting model inferences and interpretability.  Thus, \cite{hugh:hara:2013} propose to remove confounding by using the Moran basis to approximate the spatial random effect:
\[
\bfphi_j \approx \bfM\bfdelta_j,
\]
where $\bfM$ is fixed to be the first $r$ eigenvectors of $\bfP^\perp\bfA \bfP^\perp$, where $\bfP^\perp = (\bfI - \bfX(\bfX'\bfX)\inv\bfX')$ and $\bfX$ is the $n \times k$ matrix consisting of a column of ones and the economic indicators for each location, and $\bfdelta_j$ is the $r \times 1$ vector of coefficients.  Then, 
\[
\pi(\bfdelta_j | \sigma^2) \propto \left(\frac{1}{\sigma^2}\right)^{r/2} \exp\left\{-\frac{1}{2\sigma^2}\bfdelta_j'(\bfM'\bfQ\bfM)\bfdelta_j\right\},
\]
making $\bfM'\bfQ\bfM$ the precision matrix for $\bfdelta_j$.  Thus, rather than modeling the spatial random effect directly through $\bfphi_j$, we model $\bfdelta_j$.  In our analysis, we set $r = 7$ making the total number of parameters in each spatial model equal to $(k \times J) + (r \times J) + 2 = 107$. 

Finally, for the remaining parameters we use relatively non-informative prior distributions. Specifically, 
\begin{align*}
\theta &\sim \textrm{Inverse-Gamma}(a_\theta,b_\theta) \\
\m{\beta} &\sim \textrm{Normal}(\m{0},s^2 \m{I})\\
\sigma^2 &\sim \textrm{Inverse-Gamma}(a_\sigma,b_\sigma) 
\end{align*}
where $a_\theta=0.01$, $b_\theta=0.01$, $s^2= 10000$, $a_\sigma=0.01$, $b_\sigma=0.01$. We select these values for the hyperparameters so that our prior distributions are diffuse.  

\subsection{Model Selection}
We consider eight Bayesian models, comparing both spatial and non-spatial models using gamma, log-normal, normal, and Weibull likelihood distributions (see Table \ref{tab:models}; Appendix \ref{app:like} shows the parameterizations used for each likelihood distribution).  For the models using the Weibull likelihood distribution, instead of modeling the mean, $\mu_{ij}$, as a linear combination of the covariates, we model the parameter, $\lambda_{ij} = \mu_{ij}/\Gamma(1 + 1/\theta)$, as done in, for example, \cite{kalb2002}.  For the non-spatial models, we assume $\phi_{ij} \equiv 0$ for all $i$ and $j$ in equation (\ref{eq:lm}). 

Table \ref{tab:models} also defines the link function used for each model.  We chose these link functions because they are commonly-used for each likelihood and they are often used due to computational convenience \cite{mcc:nelder:1989}.  For example, the canonical link for the Gamma distribution is the inverse link, but this link is not always positive and has vertical asymptotes.  

We fit three chains (using different starting values) for each model using an adaptive Metropolis random walk sampler \citep{roberts2009} for 1500000 iterations using code written in \texttt{R} \citep{R2016}.  For memory, we saved only every tenth draw.  Of the saved draws, we kept the final 30000 as post-burn-in posterior draws.  We checked for convergence by monitoring trace plots and computing Geweke's convergence diagnostic within chains \cite{geweke1992} and Gelman-Rubin convergence diagnostic between chains \cite{gelman1992} using the \texttt{R} package \texttt{coda} \citep{coda2006}.  For all models and all chains, less than 1\% of Geweke diagnostics on parameters from all models and all chains exceeded 3 (or were below -3) and none were above (below) 3.8 (-3.8).  No Gelman-Rubin multivariate diagnostic exceeded 1.1, and the maximum univariate diagnostic value was 1.3, but almost all were below 1.1.  We also computed Monte Carlo standard errors using the \texttt{R} package \texttt{mcmcse} \citep{mcmcse2017} and these values are included with all parameter estimates in Appendix \ref{app:coef}.  

\begin{table}
\centering
\caption{Models used for comparison. Note that for the Weibull likelihood, the link is on the scale parameter, $\lambda$, instead of the mean, $\mu$.}\label{tab:models}
\begin{tabular}{cccc}
\hline
Model & Likelihood: $f(y_{ij}|\mu_{ij}, \theta)$ & Link: $g(\mu_{ij})$ & Spatial Dependence\\\hline\hline
1 & Gamma & $\log(\mu_{ij})$ & No \\
2 & Gamma & $\log(\mu_{ij})$ & Yes\\
3 & Log-normal & $\mu_{ij}$ & No\\
4 & Log-normal & $\mu_{ij}$ & Yes\\
5 & Normal & $\mu_{ij}$ & No \\
6 & Normal & $\mu_{ij}$ & Yes\\
7 & Weibull & $\log(\lambda_{ij})$ & No\\
8 & Weibull & $\log(\lambda_{ij})$ & Yes \\\hline
\end{tabular}
\end{table}

We are interested in the model that performs the best in predicting foreign aid for each sector in each district; therefore, we use the posterior predictive distribution,
\begin{equation}
f(\bm{y}_{\textrm{new}} \mid \bm{y}) = \int_\Psi  f ( \bm{y}_{\textrm{new}} \mid \m{\psi}) \, \pi( \bm{\psi|y}) \, \textrm{d}\m{\psi},\label{eq:postpred}
\end{equation}
where $\m{\psi}$ is the set of all model parameters and $\Psi$ is its parameter space, for each model to determine which model most accurately predicts foreign aid in Malawi. We consider both mean squared error (MSE) and mean absolute error (MAE) between the posterior predictive mean and observed donations.  We also compute deviance information criteria \citep[DIC;][]{spiegelhalter2002, spiegelhalter2014} for model comparison.   

\begin{table}
\centering
\small
\caption{Observed mean absolute error (MAE), mean squared error (MSE), and deviance information criteria (DIC) for the eight posed models.  The smallest number for each criterion is bolded.}\label{tab:error}
\begin{tabular}{rrccc}
\hline
\multicolumn{2}{r}{Model} & MAE &  MSE & DIC \\\hline\hline
1 & Non-spatial Gamma &10.36 & 623.88&3574.69\\
2 & Spatial Gamma & \textbf{8.93} &495.23 &\textbf{3568.18}\\
3 & Non-spatial Log-normal &10.09 &513.63 &3740.24 \\
4 & Spatial Log-normal &10.17 & 494.06&3910.73\\
5 & Non-spatial Normal &11.06 & 436.83& 4028.86\\
6 & Spatial Normal &10.83 & \textbf{420.02}&4194.91 \\
7 & Non-spatial Weibull & 13.70& 544.23& 3814.99\\
8 & Spatial Weibull &12.19 &529.77 &3860.22 \\\hline
\end{tabular}
\end{table}

Table \ref{tab:error} shows the MAE and MSE for the in-sample predictions from each of the models.  In most cases, the spatial models out predict their non-spatial counterparts; however, the improvement is modest.  Additionally, other than the spatial gamma model, the DIC is larger for the spatial models than for their non-spatial counterparts.  This indicates that while some spatial dependence is present, the additional parameters may not be worth adding for the small gains in model fit.  In this case, however, we believe a spatial model is important especially as geocoded aid data and economic data become more sophisticated and available.  That said, Model 2, or the spatial gamma model, provides the smallest MAE and the smallest DIC.  Although its DIC is comparable to the non-spatial gamma model, the MSE is much smaller.  Model 6, or the spatial normal model, has the smallest MSE.  This is due to an anomalously high value in the RPT sector (see Figure \ref{fig:donplots3} in Appendix \ref{app:preds}) that no model captures well, but the normal model captures best and therefore its MSE is less affected by the outlier.   This improvement in model fit is not seen in the MAE of the normal model because the outlier does not affect the MAE as much as the MSE.   

Table \ref{tab:error} shows that while there are some differences in model fit, all models fit the data quite similarly.  Therefore, in our analysis, we focus on the spatial gamma model because this is the model with the smallest DIC and MAE.


\section{Analysis}\label{sec:results}
In this section, we examine the parameters and fitted values of the spatial gamma model described in Section \ref{sec:model}.  In Section \ref{sec:params}, we discuss evidences for and against donor collaboration or efficient aid allocation by examining the posterior distributions of the model coefficients.  We conduct an in-depth study in Section \ref{sec:suggestions} to identify potential areas for improved aid allocation.  

\subsection{Parameter Estimates and Interpretation}\label{sec:params} 

Table \ref{tab:parameter}  shows  posterior means and standard deviations for all regression coefficients $\m{\beta}$ as well as the marginal posterior probabilities that each regression coefficient is greater than 0. We use these posterior probabilities to examine the relationship between the economic indicator and the donations within a sector.  Bolded values correspond to coefficients with posterior probability greater than 0.9 of being either positive or negative.  

We use the relationships between aid and the economic indicators to determine whether or not there is evidence of efficient aid allocation.  For example, for the Education sector, if aid is being allocated efficiently, we would expect education aid to be positively related to distance to school (greater distances to school on average indicate a greater need for education aid).  In large part, though, there is not evidence that the economic indicators are related to districts' aid per person within a sector at all.  Many of the coefficient posterior distributions envelop 0 (those non-bolded coefficients).  In fact, two sectors -- Governance and Water, Sanitation, and Irrigation (WSI) -- appear to not be related to any of the economic indicators.  For Governance, this isn't surprising as more than 50\% of the donations from that sector were given at the country or region level (see Figure \ref{fig:country} and \ref{fig:region}), and we assigned these projects equally per person to each of the districts within the country or region, respectively. For WSI, however, this is not the case and is more likely explained by WSI aid not being influenced by the economic indicators.  

There are, however, several coefficients that provide evidence of efficient aid allocation.  For example, mean land cultivated per household has a 0.95 posterior probability of being positively related to agriculture aid.  This makes sense since more land cultivated indicates more agricultural areas and thus more need for agricultural aid.  Additionally, distance to school is positively related to RD aid (0.98 posterior probability), as expected, since districts where the people are farther from a school tend to be more rural and thus would benefit from RD aid. 

\begin{landscape}
\begin{table}
\small
\centering
\caption{Summary of the posterior distributions of the coefficients of the spatial gamma model.  `$P(>0)$' means the posterior probability that the corresponding coefficient is greater than 0.  Coefficients (other than intercepts) that have $>$90\% posterior probability of being either positive or negative are bolded. RD$=$Integrated Rural Development; RPT$=$Roads, Public Works, \& Transportation; WSI$=$Water, Sanitation, Irrigation.}\label{tab:parameter}
\begin{tabular}{rccccccc}
Sector & \emph{Agriculture} & \emph{Education} & \emph{Governance} & \emph{Health} & \emph{RD} & \emph{RPT} & \emph{WSI}\\\hline
Covariate & mean (sd) &mean (sd) &mean (sd) &mean (sd)&mean (sd)&mean (sd) &mean (sd) \\[-.03in]
& {\footnotesize $P(>0)$} & {\footnotesize $P(>0)$} & {\footnotesize $P(>0)$} & {\footnotesize $P(>0)$}  & {\footnotesize $P(>0)$}  & {\footnotesize $P(>0)$}  & {\footnotesize $P(>0)$} \\\hline\hline
Intercept &3.55 (0.08)&3.5 (0.08)&4.21 (0.03)&4.46 (0.03)&2.69 (0.12)&2.54 (0.17)&2.76 (0.12)\\[-.03in]
&{\footnotesize { 1.00 }}&{\footnotesize { 1.00 }}&{\footnotesize { 1.00 }}&{\footnotesize { 1.00 }}&{\footnotesize { 1.00 }}&{\footnotesize { 1.00 }}&{\footnotesize { 1.00 }}\\
Poverty &-0.08 (0.11)&0.09 (0.12)&-0.02 (0.06)&\textbf{ 0.07 } (0.04)&\textbf{ -0.29 } (0.18)&\textbf{ 0.34 } (0.22)&-0.07 (0.21)\\[-.03in]
&{\footnotesize 0.22 }&{\footnotesize 0.80 }&{\footnotesize 0.37 }&{\footnotesize \textbf{ 0.97 }}&{\footnotesize \textbf{ 0.04 }}&{\footnotesize \textbf{ 0.94 }}&{\footnotesize 0.37 }\\
Dist to School &0.00 (0.13)&-0.16 (0.14)&0.03 (0.07)&-0.02 (0.05)&\textbf{ 0.42 } (0.2)&-0.11 (0.24)&0.21 (0.23)\\[-.03in]
&{\footnotesize 0.49 }&{\footnotesize 0.13 }&{\footnotesize 0.66 }&{\footnotesize 0.31 }&{\footnotesize \textbf{ 0.98 }}&{\footnotesize 0.31 }&{\footnotesize 0.81 }\\
All Injured &\textbf{ -0.17 } (0.11)&\textbf{ 0.17 } (0.09)&0.00 (0.05)&0.01 (0.04)&-0.14 (0.15)&\textbf{ 0.59 } (0.16)&0.06 (0.15)\\[-.03in]
&{\footnotesize \textbf{ 0.06 }}&{\footnotesize \textbf{ 0.97 }}&{\footnotesize 0.52 }&{\footnotesize 0.65 }&{\footnotesize 0.17 }&{\footnotesize \textbf{ 1.00 }}&{\footnotesize 0.66 }\\
Food \% &0.1 (0.13)&0.04 (0.17)&-0.01 (0.05)&0.00 (0.04)&0.06 (0.15)&\textbf{ -0.68 } (0.17)&-0.15 (0.17)\\[-.03in]
&{\footnotesize 0.81 }&{\footnotesize 0.5 }&{\footnotesize 0.42 }&{\footnotesize 0.43 }&{\footnotesize 0.64 }&{\footnotesize \textbf{ 0 }}&{\footnotesize 0.16 }\\
Land Cultivated &\textbf{ 0.13 } (0.08)&-0.02 (0.08)&-0.02 (0.04)&\textbf{ -0.11 } (0.03)&-0.08 (0.12)&\textbf{ -0.52 } (0.12)&-0.03 (0.12)\\[-.03in]
&{\footnotesize \textbf{ 0.95 }}&{\footnotesize 0.43 }&{\footnotesize 0.31 }&{\footnotesize \textbf{ 0.00 }}&{\footnotesize 0.23 }&{\footnotesize \textbf{ 0.00 }}&{\footnotesize 0.38 }\\
Electric \% &\textbf{ 0.20 } (0.07)&-0.03 (0.08)&0.03 (0.04)&\textbf{ 0.04 } (0.03)&0.08 (0.09)&\textbf{ 0.94 } (0.12)&-0.15 (0.17)\\[-.03in]
&{\footnotesize \textbf{ 0.99 }}&{\footnotesize 0.37 }&{\footnotesize 0.80 }&{\footnotesize \textbf{ 0.93 }}&{\footnotesize 0.81 }&{\footnotesize \textbf{ 1.00 }}&{\footnotesize 0.19 }\\
Population &-0.05 (0.10)&-0.06 (0.09)&0.04 (0.04)&-0.02 (0.03)&-0.11 (0.10)&\textbf{ 0.39 } (0.11)&0.09 (0.13)\\[-.03in]
&{\footnotesize 0.34 }&{\footnotesize 0.28 }&{\footnotesize 0.85 }&{\footnotesize 0.25 }&{\footnotesize 0.13 }&{\footnotesize \textbf{ 1.00 }}&{\footnotesize 0.78 }\\\hline
\end{tabular}
\end{table}
\end{landscape}

In contrast, poverty has a posterior probability of 0.96 of being negatively related to RD aid.  This means that as the proportion of the population living below the poverty line increases, the amount of RD aid decreases, contradicting what we would expect for efficient aid allocation.  Also, number injured is positively related to education aid (posterior probability of 0.97), but there is not an intuitive reason for why this would make sense.  A natural economic indicator for education would be distance to school, but this relationship is weaker (and negative -- the opposite of what we would expect).

Several economic indicator variables are related to the RPT sector.  This is likely due to a large outlier in this sector, visible in Figure \ref{fig:hist}.  That said, the positive relationship between poverty proportion, electric lighting proportion, and population with RPT aid, and the negative relationship between food as a percent of expenditures and mean land cultivated per household with RPT aid indicate that RPT aid is primarily being given to highly populated and less rural areas.  This does not match intuition.  We would expect more \emph{total} money would be needed in populated and urban areas, but that more money \emph{per person} would be needed in rural areas for building and maintaining roads, public works, and transportation.

Because the eight models fit similarly (see Table \ref{tab:error}), 
we provide coefficient summaries for all models and all coefficients in Appendix \ref{app:coef}.   
As expected, the coefficients follow similar patterns across models and most of the time, models find the same variables to have high posterior probabilities of being different from zero.  

Obviously, these interpretations of coefficients are conditional on the values of the other covariates and multicollinearity among the covariates could introduce changes in the values of the coefficients when some covariates are left out of the model (thus changing interpretations).  However, there is not much collinearity between the economic indicators, the highest absolute value correlation is 0.3, with most absolute value correlations less than 0.15.  Table \ref{tab:vif} provides variance inflation factors (VIF) for each economic indicator and scaled generalized variance inflation factors  \citep[GVIF;][]{fox1992} which account for the interaction between sector and each economic indicator.  All VIF's are below the commonly-used rule of thumb of five to indicate concerning multicollinearity \cite[see, e.g.,][]{dobson2008}. That said, interpretations of the coefficients must be made with care.  

\begin{table}[b]
\centering
\caption{Variance inflation factors (VIF) and scaled Generalized VIF's (GVIF) for economic indicator variables.}\label{tab:vif}
\begin{tabular}{ccc}
\hline
Covariate & VIF & GVIF$^{1/14}$\\\hline
Poverty & 2.49 & 1.58\\
Dist to School & 3.54 & 1.88\\
All Injured & 2.36 & 1.54\\
Food \% & 1.94 & 1.39\\
Land Cultivated & 1.40 & 1.18 \\
Electric \% & 1.21 & 1.10\\
Population & 1.11 & 1.05\\\hline
\end{tabular}
\end{table}

\begin{figure}
 \begin{center}
     \begin{subfigure}[h]{\textwidth}
  \includegraphics[width=.32\textwidth]{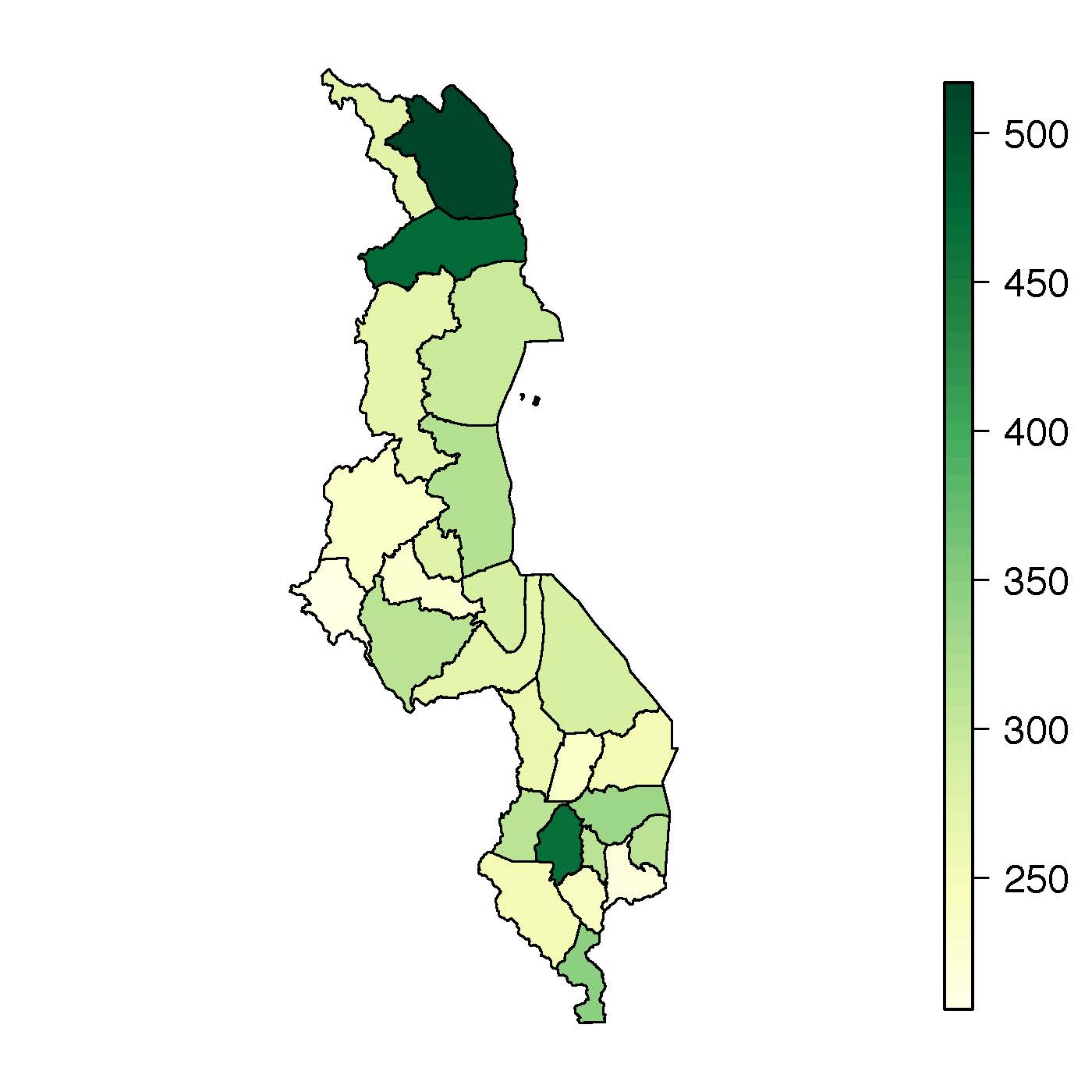}
  \includegraphics[width=.32\textwidth]{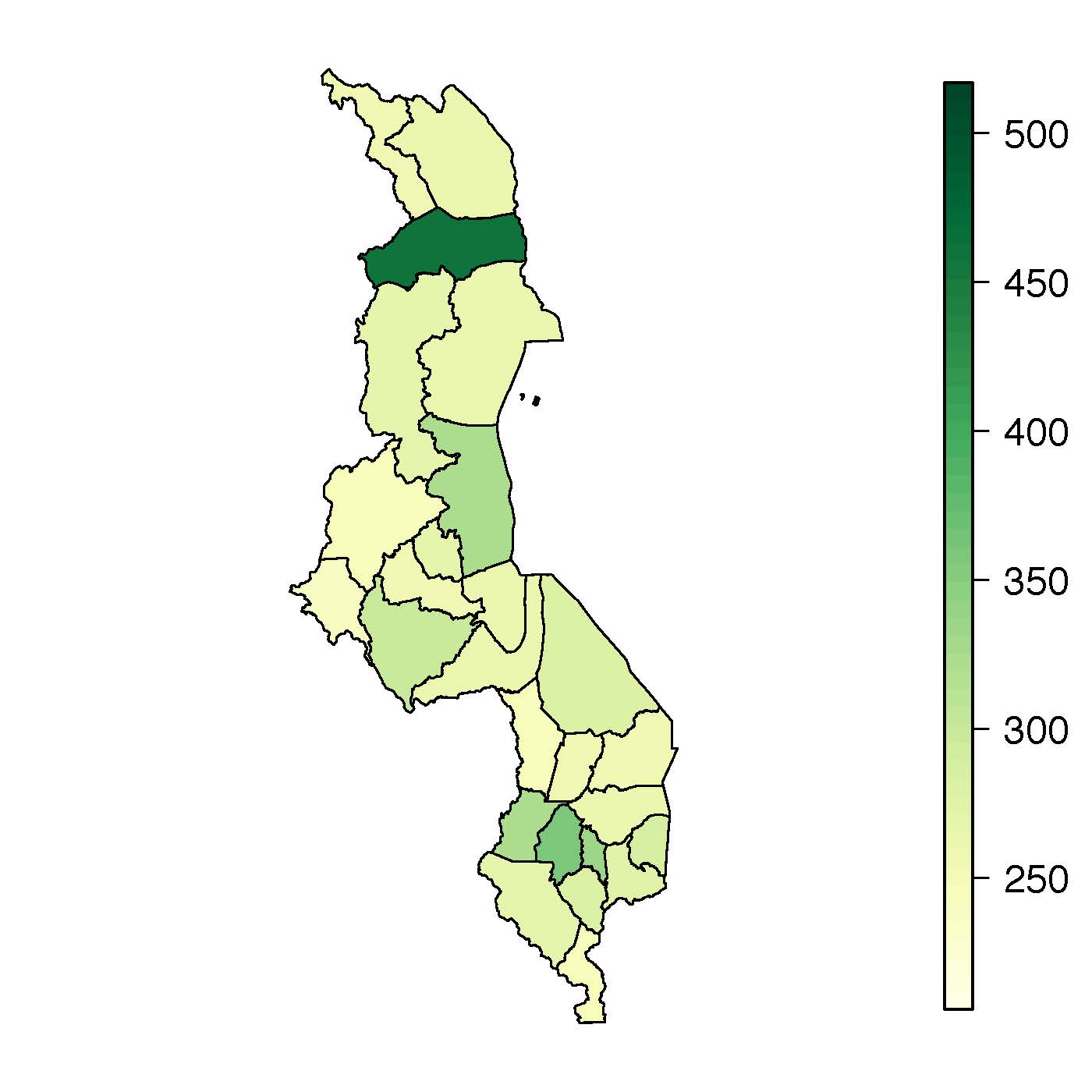}
  \includegraphics[width=.32\textwidth]{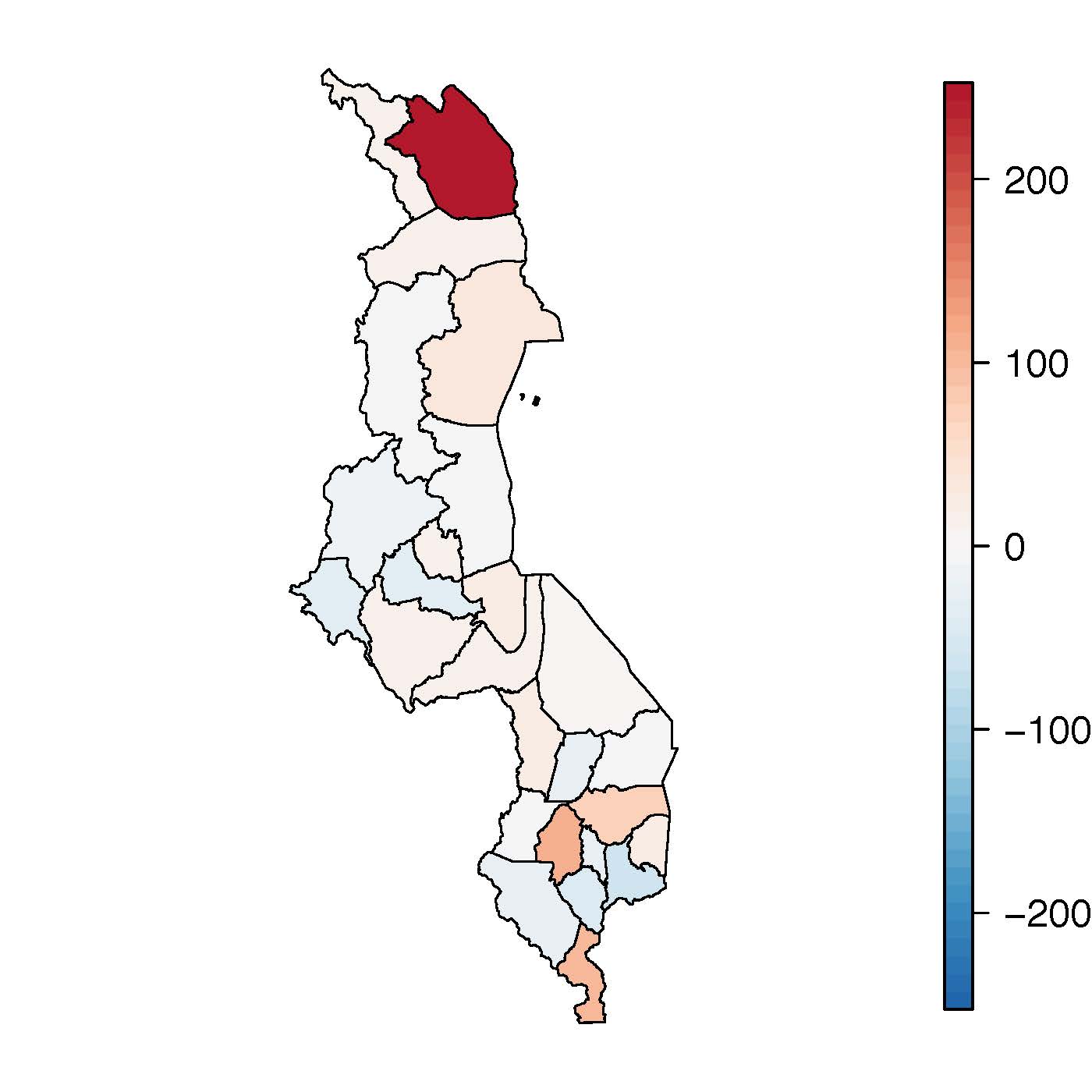}
  \vspace{-3mm}
 \subcaption{Total Donation}\label{fig:total}
 \end{subfigure}
    \begin{subfigure}[h]{\textwidth}
  \includegraphics[width=.32\textwidth]{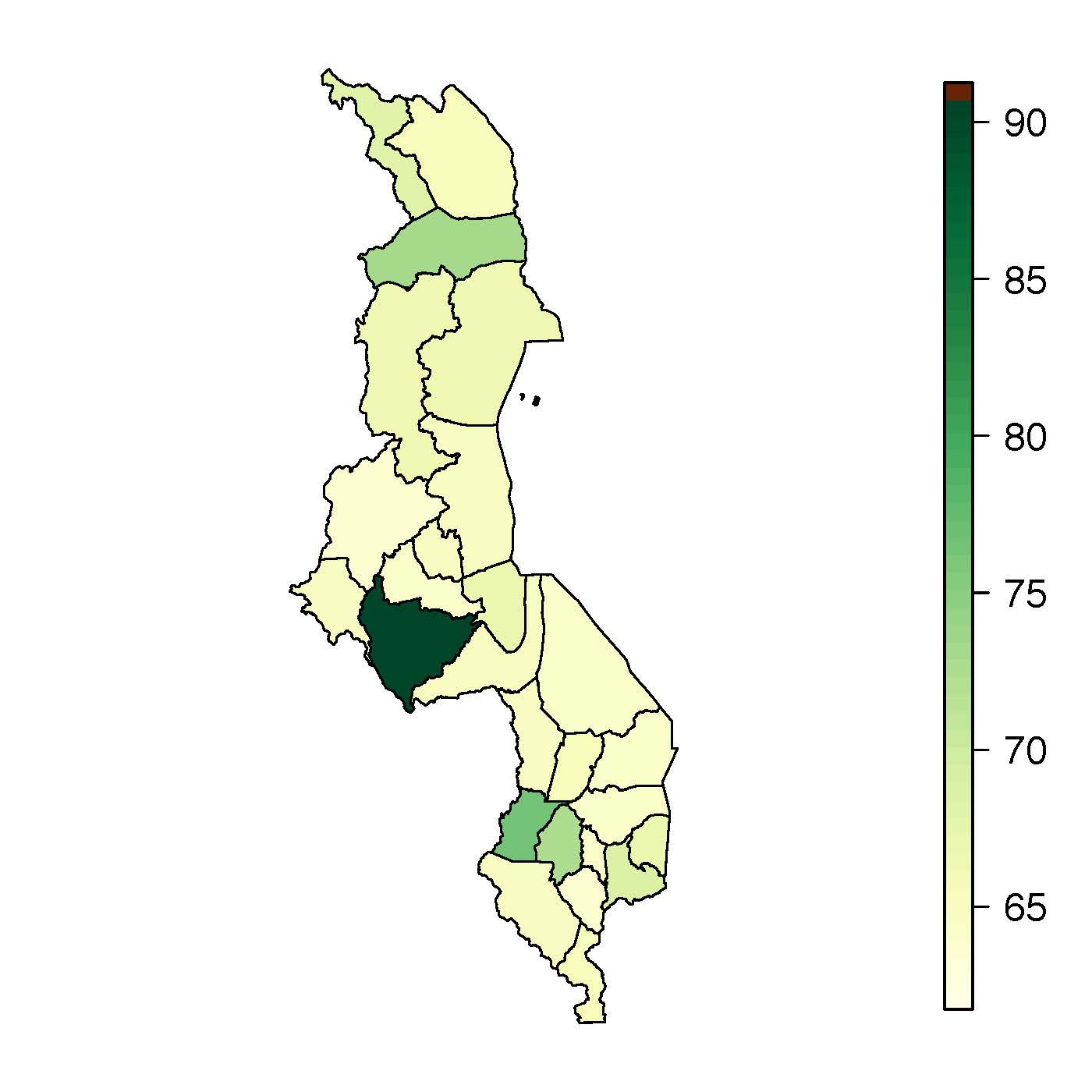}
  \includegraphics[width=.32\textwidth]{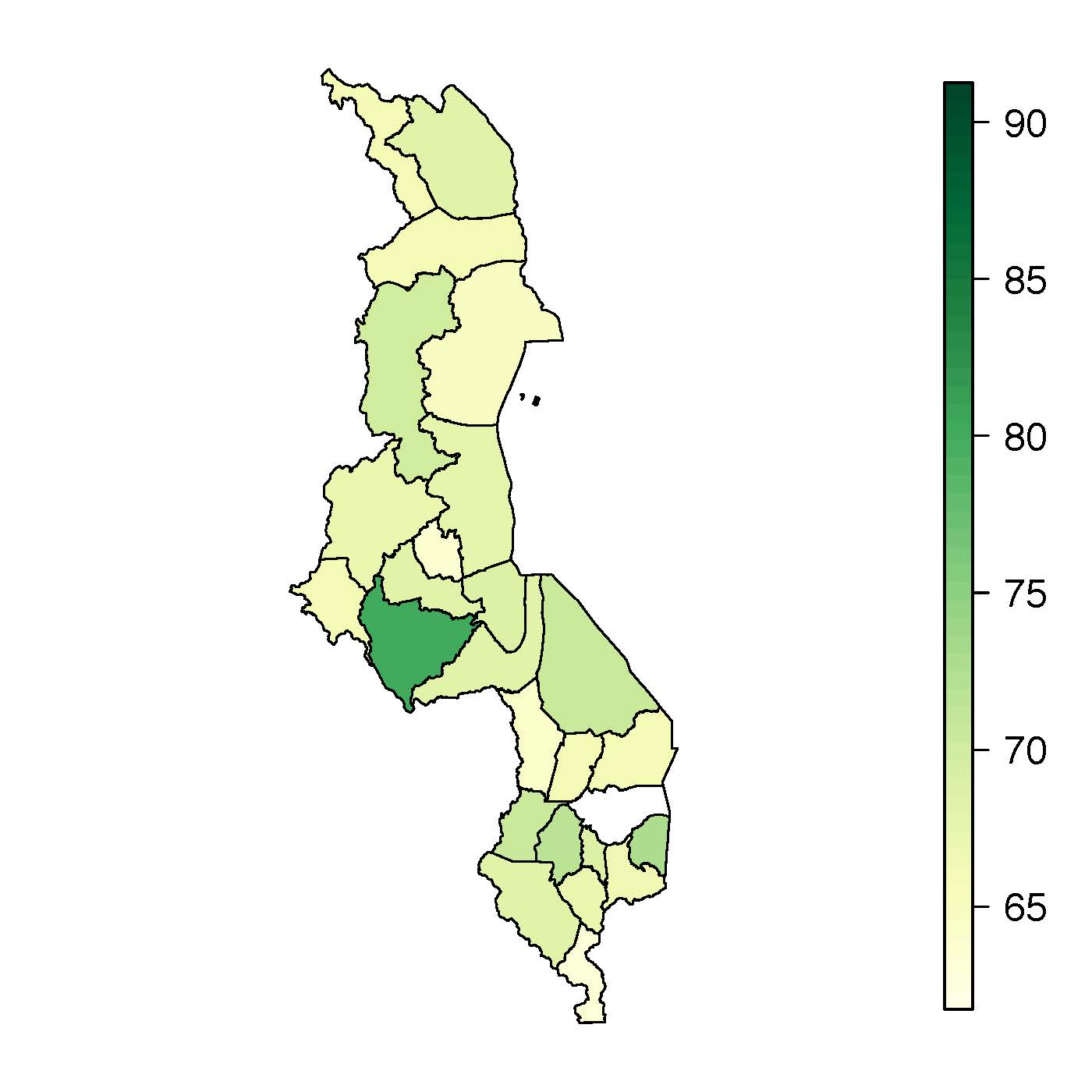}
  \includegraphics[width=.32\textwidth]{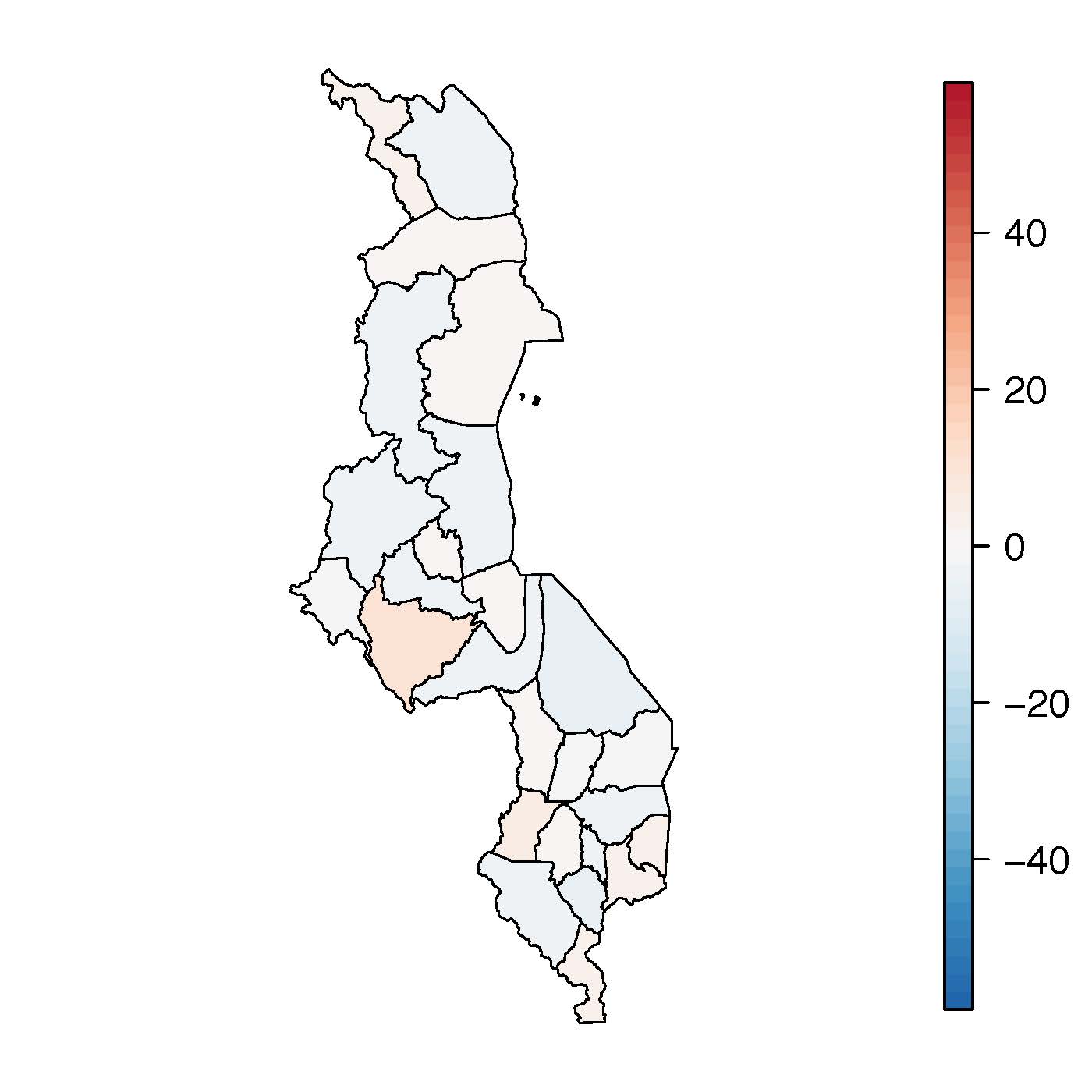}
  \vspace{-3mm}
 \subcaption{Governance}\label{fig:gov}
  \end{subfigure}
     \begin{subfigure}[h]{\textwidth}
  \includegraphics[width=.32\textwidth]{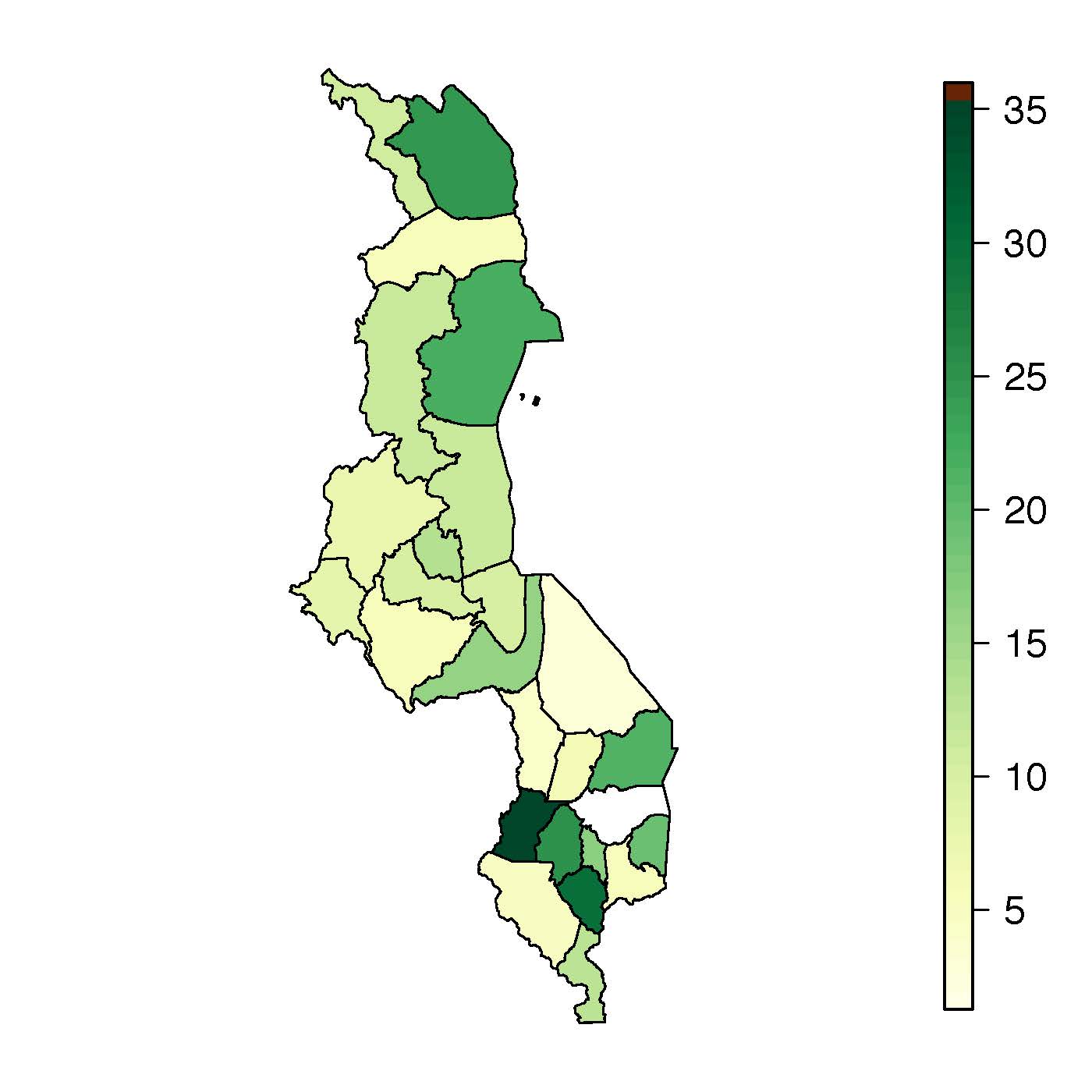}
  \includegraphics[width=.32\textwidth]{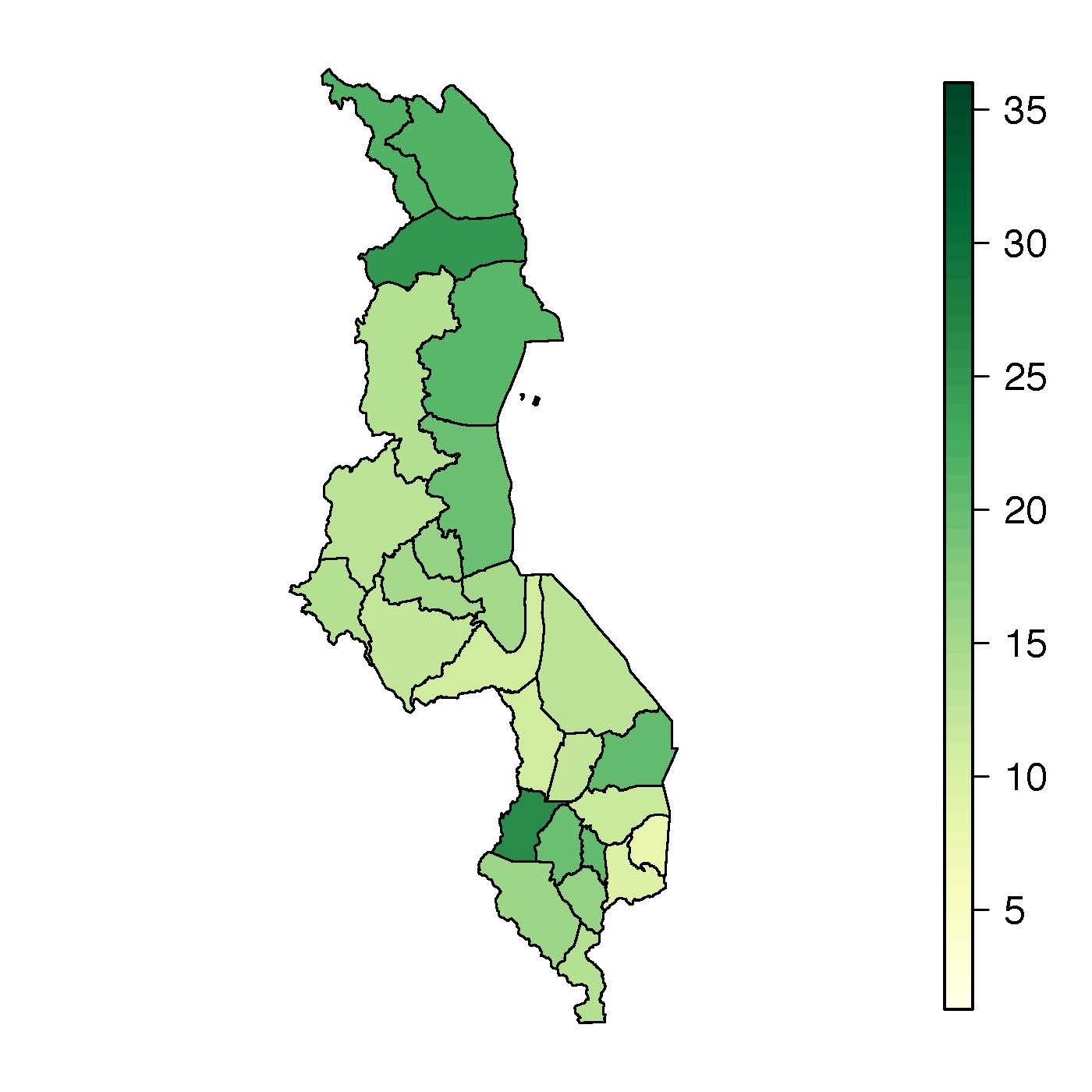}
  \includegraphics[width=.32\textwidth]{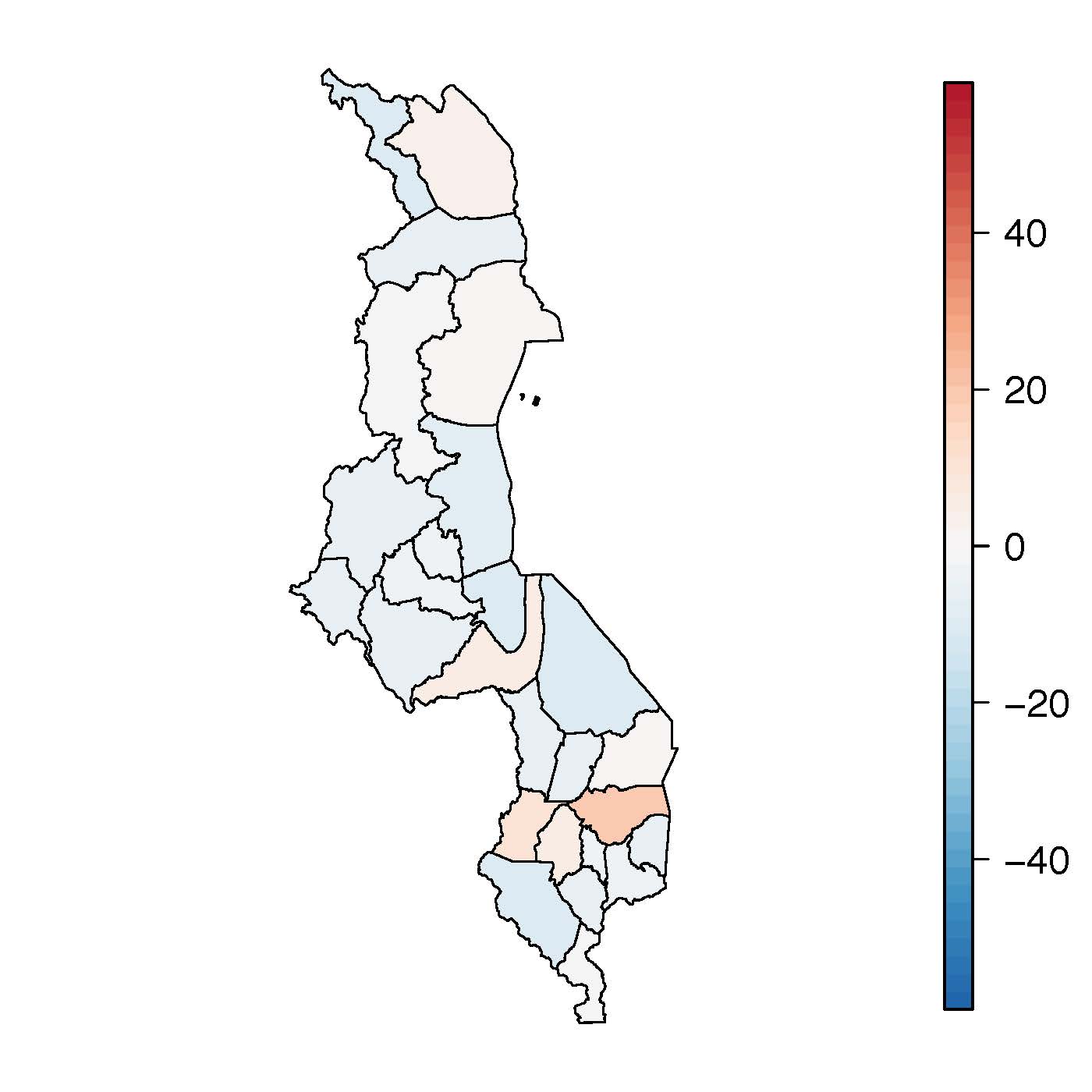}
  \vspace{-3mm}
 \subcaption{Rural Development}\label{fig:rural}
  \end{subfigure}
    \caption{Observed (left), predicted (center), and residual (right) donations plots for total donations (a) and the Governance (b) and RD (c) sectors. All values are in USD.}\label{fig:donplots1}
 \end{center}
\end{figure}

Figure \ref{fig:donplots1} shows plots of observed, posterior predictive means, and the residuals of total donations, and donations for Governance and RD.  Plots for the remaining sectors are in Appendix \ref{app:preds} in Figures \ref{fig:donplots2} and \ref{fig:donplots3}.  


\subsection{More Efficient Aid Allocation}\label{sec:suggestions}
Although we do not expect the economic indicators to be perfect drivers of aid allocation, they can provide direction for better aid allocation.  In this section, we consider the scenario where aid is allocated according to a more expected or ideal relationship with specific economic indicators.  We call this scenario our ``efficient scenario.''  By doing this, we can first examine the potential for our model to capture a ``true'' relationship between aid and economic indicators.  Second, and more important for donors as they identify locations to send aid, is to identify locations where more or less aid is needed within each sector.  

To accomplish this task, we select a single economic indicator for each sector and identify its expected relationship to aid, shown in Table \ref{tab:idealcovs}.  We then use the posterior distributions from the fitted model of Section \ref{sec:params}, but adjust the coefficients for these selected economic indicators.  Specifically, we set the mean of the corresponding coefficient of each economic indicator of interest to be three posterior standard deviations above (or below for a negative relationship) zero.  For example, the adjusted mean of the coefficient corresponding to Land Cultivated for Agriculture is set to 0.24 (see Table \ref{tab:parameter}), while the other coefficients for the Agriculture sector remain at the original posterior mean.  Then, the adjusted posterior distribution on the coefficients is
\begin{equation}
\bfbeta^* \sim N(\bar{\bfbeta}^*, \bfSigma(\bfbeta)),\label{eq:adjpost}
\end{equation}
where  $\bar{\bfbeta}^*$ is the adjusted posterior mean of $\bfbeta$ as described above and $\bfSigma(\bfbeta)$ is the posterior variance of $\bfbeta$ from the fitted spatial gamma model in Section \ref{sec:params}.  The posterior correlation among the $\bfbeta$'s was quite small, and thus $\bfSigma(\bfbeta)$ could also be a diagonal matrix with the marginal variances of each coefficient on the diagonal.  

\begin{table}
\begin{center}
\begin{tabular}{ccc}
\hline
Sector & Economic Indicator & Expected Relationship\\\hline
Agriculture & Land Cultivated & Positive\\
Education & Dist to School & Positive \\
Governance & Poverty & Positive\\
Health & All Injured & Positive \\
RD & Electric \% & Negative\\
RPT & Electric \% & Negative \\
WSI & Land Cultivated & Positive \\\hline
\end{tabular}
\caption{Selected economic indicator for each sector and the corresponding direction of the relationship.}\label{tab:idealcovs}
\end{center}
\end{table}

We then obtain adjusted posterior predictive distributions on aid per person for each sector using equation (\ref{eq:postpred}), but with the adjusted posterior distribution of $\bfbeta$. The computational algorithm is as follows: 
\begin{enumerate}
\item Draw a value of $\bfbeta^*$ from equation (\ref{eq:adjpost}).
\item Use a draw of $\bfphi$ and $\theta$ from the posterior draws obtained in Section \ref{sec:params}.
\item Draw a value of the adjusted aid allocation, $y^*_{ij}$, for each $i$ and $j$ from the gamma likelihood, conditional on the values in steps 1 and 2.
\item Scale $y^*_{ij}$ so that the new sector totals are the same as the data sector totals: $\sum_{i = 1}^{26} y^*_{ij} = \sum_{i =1}^{26} y_{ij}$.  (This requires that the total money within a sector does not change, only its distribution across the districts.) 
\item Repeat for all post-burn-in posterior draws.
\end{enumerate}
We can then use these draws from the adjusted posterior predictive distributions to make inferences on aid allocation.  

To learn about the ability of our model to identify a ``true'' relationship, we refit the spatial gamma model to 26 randomly selected draws from the adjusted posterior predictive distribution.  In this way we are making use of frequentist sampling distribution ideas to compute a power-type diagnostic.  For these 26 randomly selected draws, using a 90\% posterior probability of being greater than zero (or less than zero for expected negative relationships) cut-off, we would have identified a relationship {85\%} of the time.  This is a bit lower than 90\%, but this may be due to re-scaling the data to keep total donation amounts the same.  Notably, however, we never found a relationship in the opposite direction; i.e., a parameter that was truly positive (negative) never had a 90\% posterior probability of being negative (positive).  Recall that RPT was positively related to electric lighting proportion -- the opposite of what we would hope to see.  This simulation study provides more evidence that if RPT aid were actually negatively related to electric lighting proportion, we would not have found this strong positive relationship.

\begin{figure}[H]
\begin{center}
\includegraphics[width=6in]{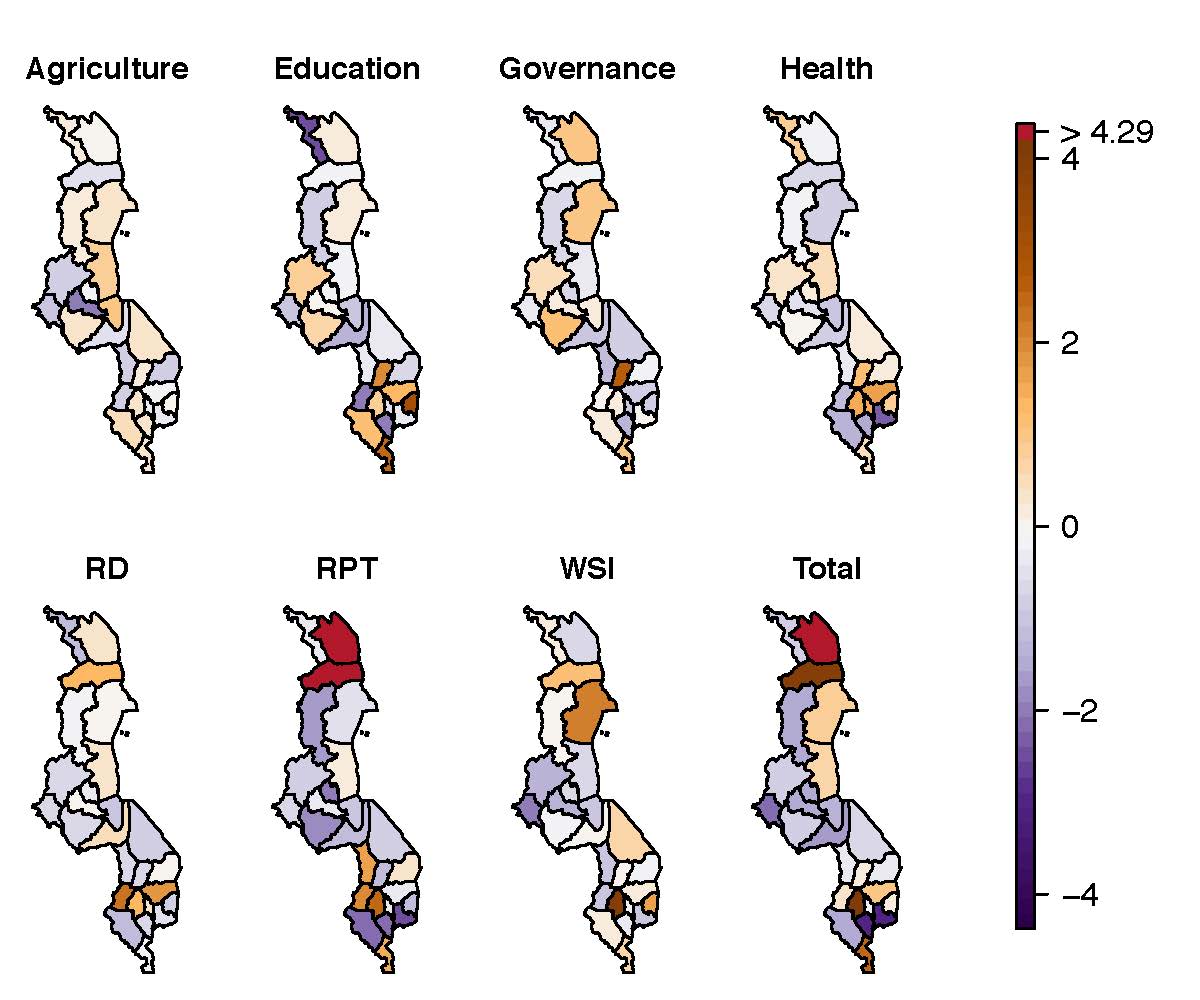}
\caption{Standardized residual maps between the observed and efficient scenario donations per person for each sector and for all sectors combined (bottom right).}\label{fig:adjresids}
\end{center}
\end{figure}

Finally, we compare the adjusted posterior predictive distributions of donation amounts to the observed donation amounts.  Figure \ref{fig:adjresids} shows standardized residual maps; specifically, for district $i$ and sector $j$,
\[
r_{ij} = \frac{y_{ij} - \bar{y}^*_{ij}}{SD(y^*_{ij})}, 
\]
where $y_{ij}$ is the observed aid per person, $\bar{y}^*_{ij}$ is the mean of the adjusted posterior predictive distribution of aid per person, and $SD(y^*_{ij})$ is the standard deviation of the adjusted posterior predictive distribution of aid per person.  By standardizing this way, we can compare differences between the predictive means and the observed values relative to the variation of the predictive distributions, so that a large difference between the observed and predictive means is either dampened or enhanced by a large or small standard deviation, respectively.  In Figure \ref{fig:adjresids}, darker orange values indicate that more aid is going to these locations than our efficient scenario suggests, while dark purple values indicate that less aid is going to these locations than our efficient scenario suggests.    
Looking at total donations per person within each district, many of the districts that indicate more need for aid are on the western side of Malawi.  

\section{Conclusion}\label{sec:conclusion}

Foreign aid is important for many countries' economies and its allocation can be even more important to both the donors and citizens within the receiving country.  But efficient allocation is a difficult problem and identifying whether or not it is occurring is difficult too.  In this paper, we made use of a unique dataset and examined how aid allocation within Malawi is related to various economic indicators within seven donation sectors.  We found some evidence of efficient allocation, but on the whole, there is not evidence that aid is being allocated according to economic indicators. 
Additionally, because it is so hard to efficiently allocate donations, there is a need for identifying ways to help.  We also used this model to develop a more efficient aid allocation according to a specific subset of the economic indicators and provided maps of adjustments to current aid allocation. 

Although most countries do not have databases with similar location-specific information on donations, these methods can be applied to examine efficiency of cross-national aid allocation. And as more subnational geocoded aid data become available, these methods will become more relevant. 


\bibliography{refs}

\appendix

\section{Likelihood Parameterizations}\label{app:like}
  In this appendix we include the parameterization used for each likelihood distribution in Models 1--8.  For brevity, we drop subscript notation so that below, $y_{ij} = y$, $\mu_{ij} = \mu$, $\bfx_{ij} = \bfx$, and $\bfbeta_j = \bfbeta$.  
  
  \begin{description}
  \item[Gamma]
  \[
  f(y|\mu, \theta) = \left(\frac{\mu}{\theta}\right)^{\mu^2/\theta} \frac{1}{\Gamma(\mu^2/\theta)}\,\, y ^{(\mu^2/\theta) - 1} \,\exp\{-y \mu/\theta\},
  \]
  where $E(y|\mu, \theta) = \mu$ and $\mbox{Var}(y|\mu, \theta)= \theta$.  
  \item[Lognormal]
  \[
  f(y|\mu, \theta) = \frac{1}{\sqrt{2\pi\theta} y} \exp\left\{-\frac{1}{2\theta} (\log y - \mu)^2\right\},
  \]
 where $E(\log y|\mu, \theta) = \mu$ and $\mbox{Var}(\log y | \mu, \theta) = \theta$. 
  \item[Normal]
  \[
   f(y|\mu, \theta) = \frac{1}{\sqrt{2\pi\theta}} \exp\left\{-\frac{1}{2\theta} (y - \mu)^2\right\},
   \]
   where $E(y|\mu, \theta) = \mu$ and $\mbox{Var}(y|\mu, \theta)= \theta$. 
  \item[Weibull] 
  \[
  f(y|\lambda, \theta) = \frac{\theta}{\lambda} \left(\frac{y}{\lambda}\right)^{\theta - 1} \exp\left\{-\left(\frac{y}{\lambda}\right)^\theta\right\},
  \]
  where  $E(y|\lambda, \theta) = \lambda \Gamma(1 + 1/\theta)$ and $\mbox{Var}(y|\lambda, \theta)= \lambda^2 (\Gamma(1 + 2/\theta) - \Gamma(1+1/\theta)^2)$.  Note that this parameterization corresponds to the Weibull regression model found in \cite{kalb2002} as follows.  As in our models, let $\log \lambda = \bfx'\bfbeta = \beta_0 + \bfx_{1}'\bfbeta_1$, where $\beta_0$ corresponds to the intercept, $\bfx_{1}$ corresponds to the vector of covariates without the intercept, and $\bfbeta_1$ corresponds to the remaining vector of coefficients (all the coefficients without the intercept).  Then, the pdf in (2.8) of \cite{kalb2002} can be written by letting $\theta = \gamma$, $\exp\{-\beta_0\} = \lambda$, and $-\bfbeta_1\theta = \bfbeta$ where the parameters on the right correspond to the notation used in \cite{kalb2002} and the parameters on the left correspond to the parameters used in our model.    
%
  \end{description}

\section{Predicted Values}\label{app:preds}
\begin{figure}[H]
 \begin{center}
   \begin{subfigure}[h]{\textwidth}
  \centering
  \includegraphics[width=.32\textwidth]{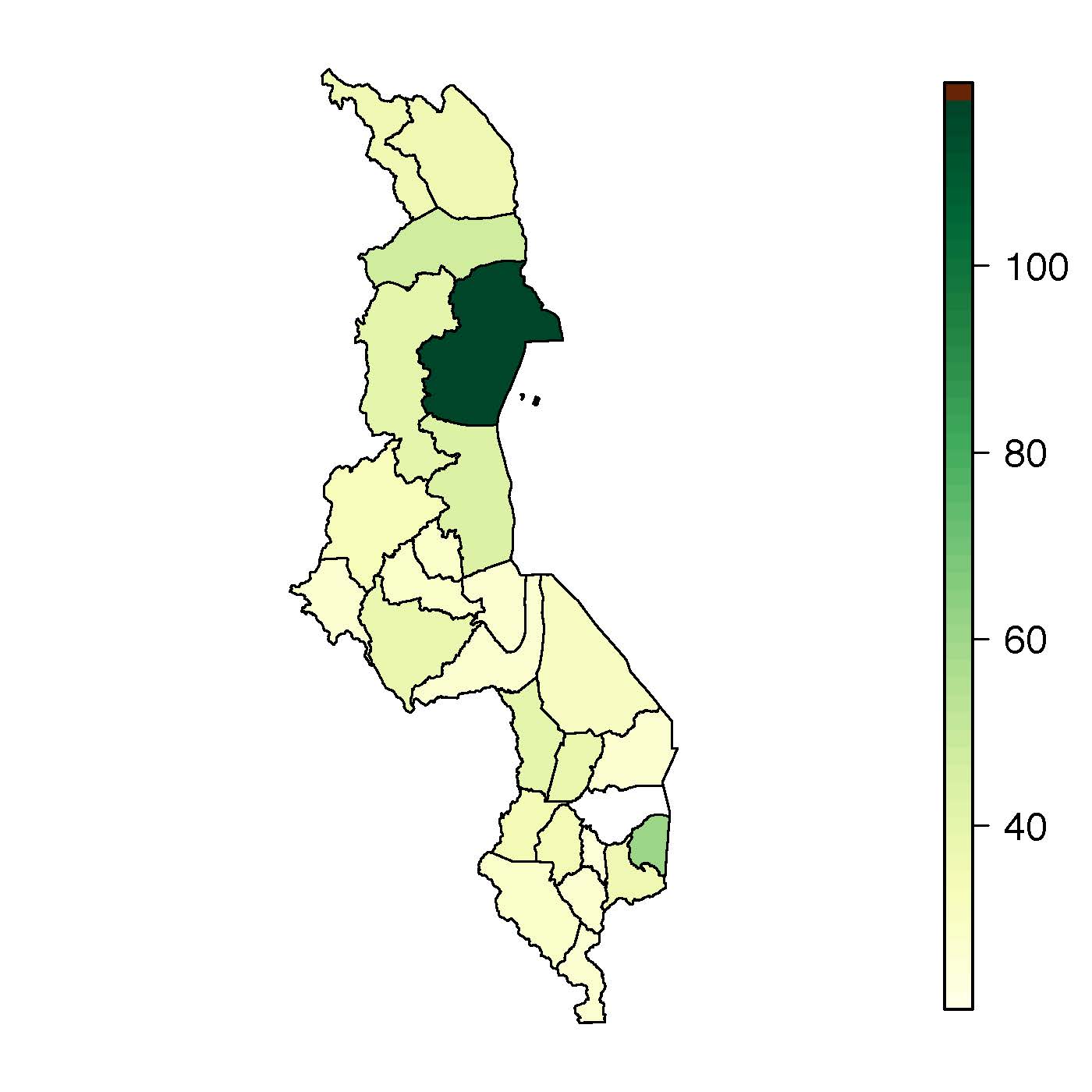}
  \includegraphics[width=.32\textwidth]{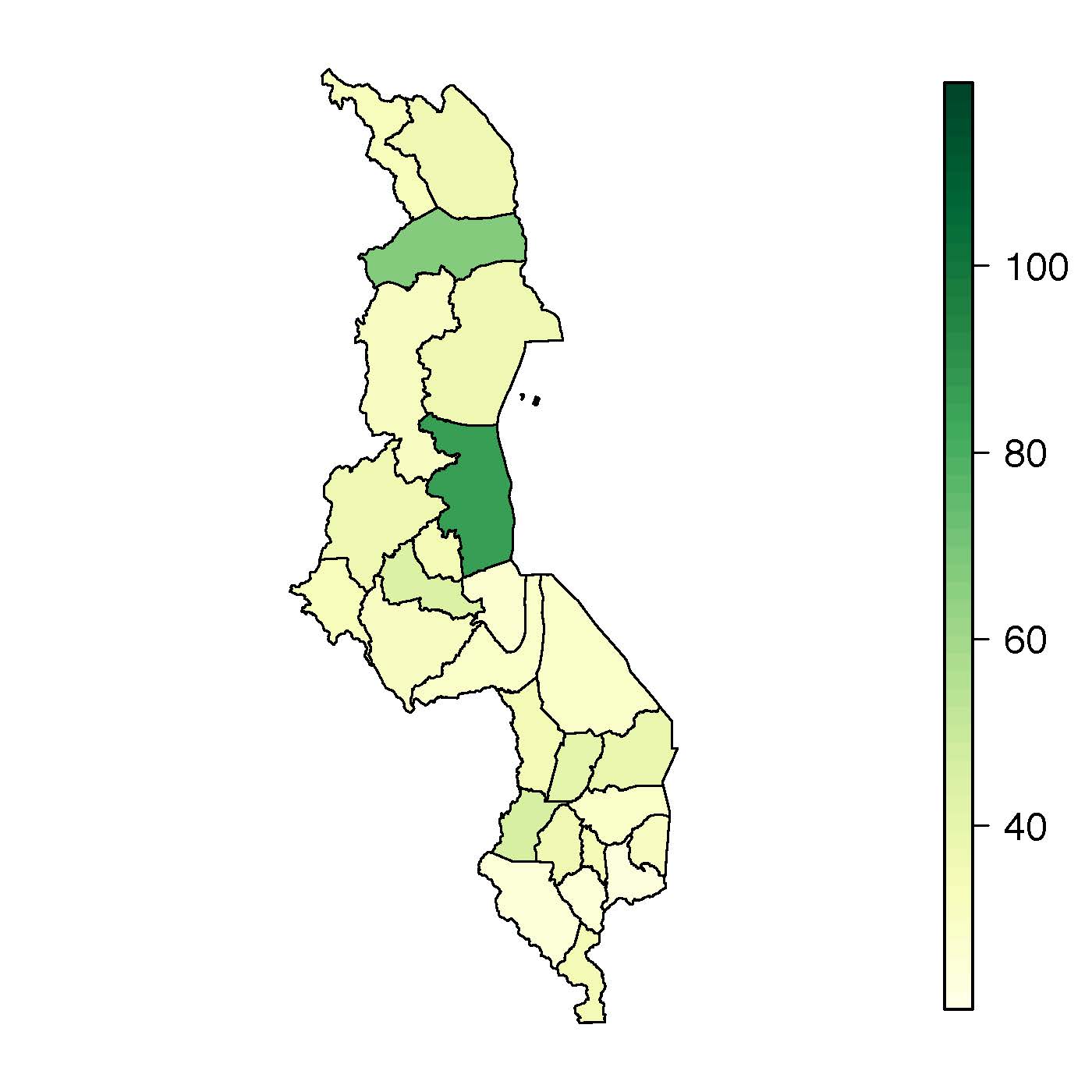}
  \includegraphics[width=.32\textwidth]{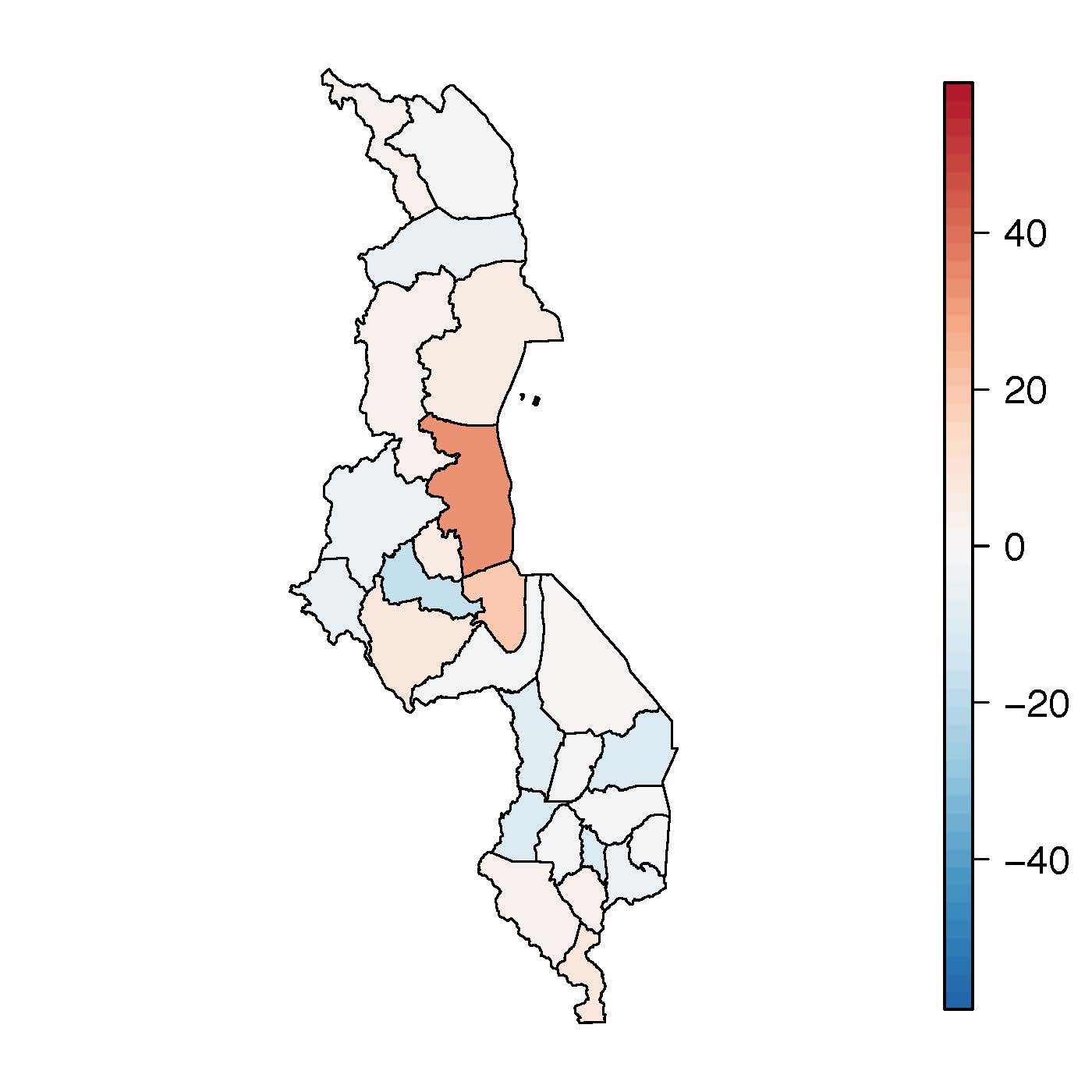}
  \vspace{-3mm}
 \subcaption{Agriculture}\label{fig:ag}
  \end{subfigure}
   \begin{subfigure}[h]{\textwidth}
  \centering
  \includegraphics[width=.32\textwidth]{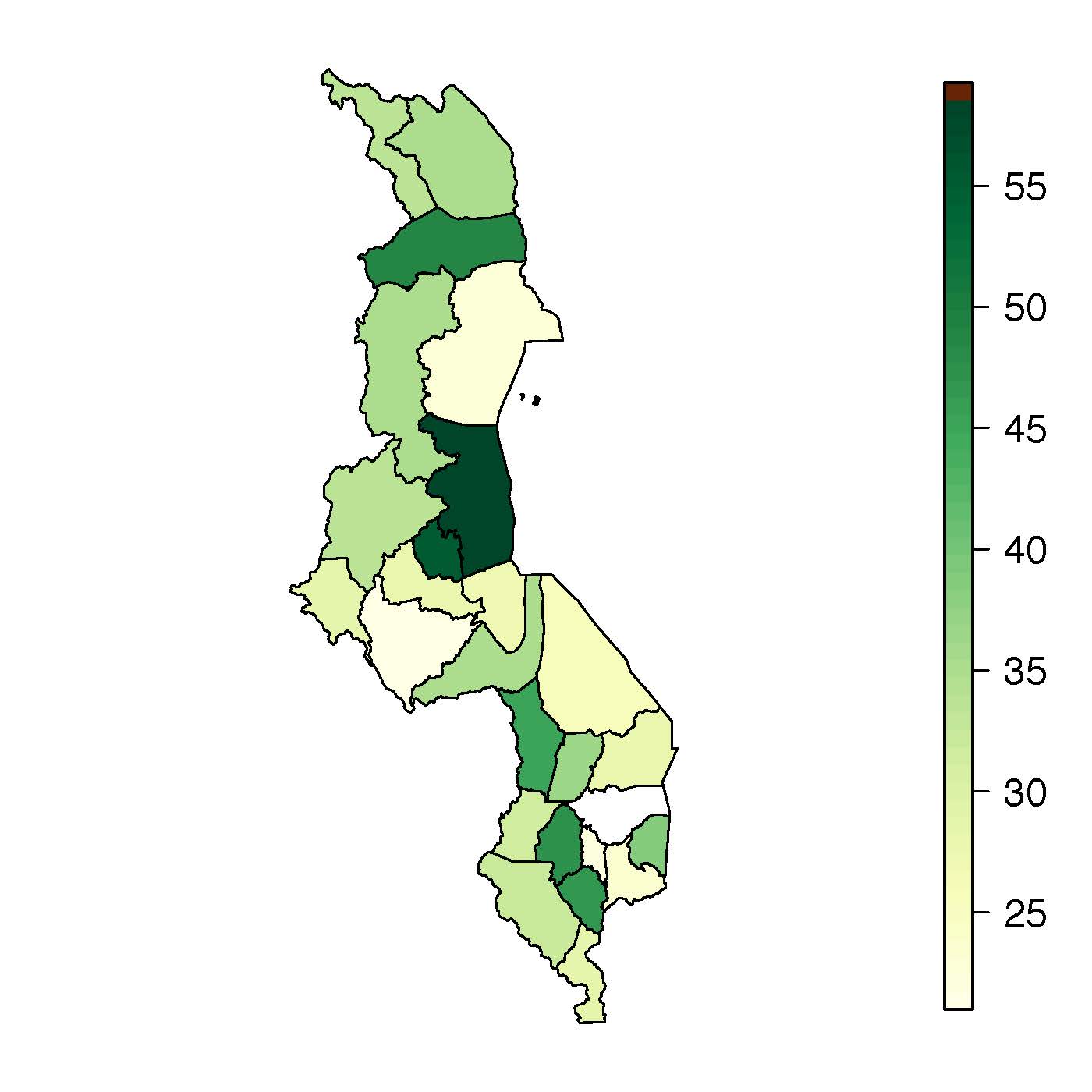}
  \includegraphics[width=.32\textwidth]{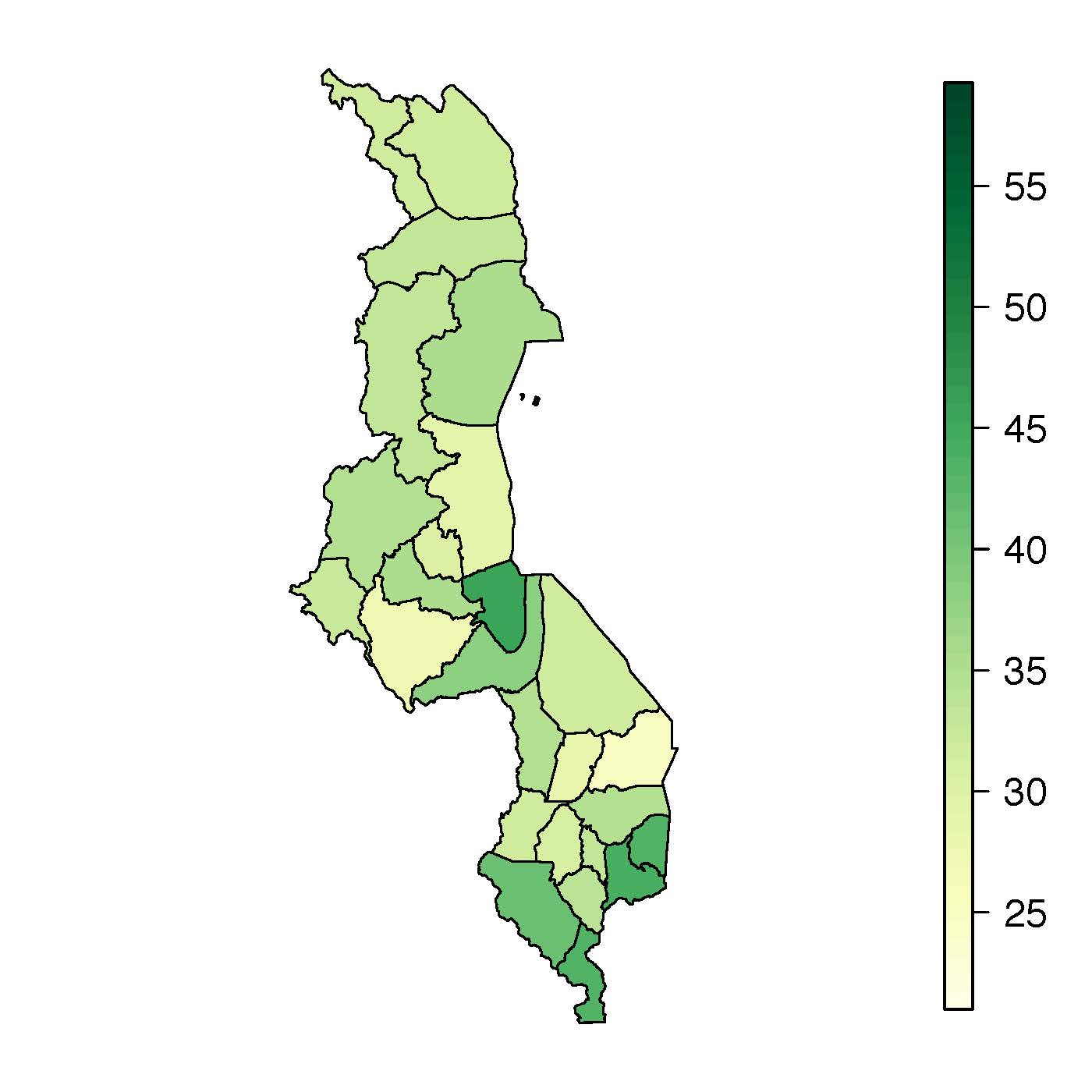}
  \includegraphics[width=.32\textwidth]{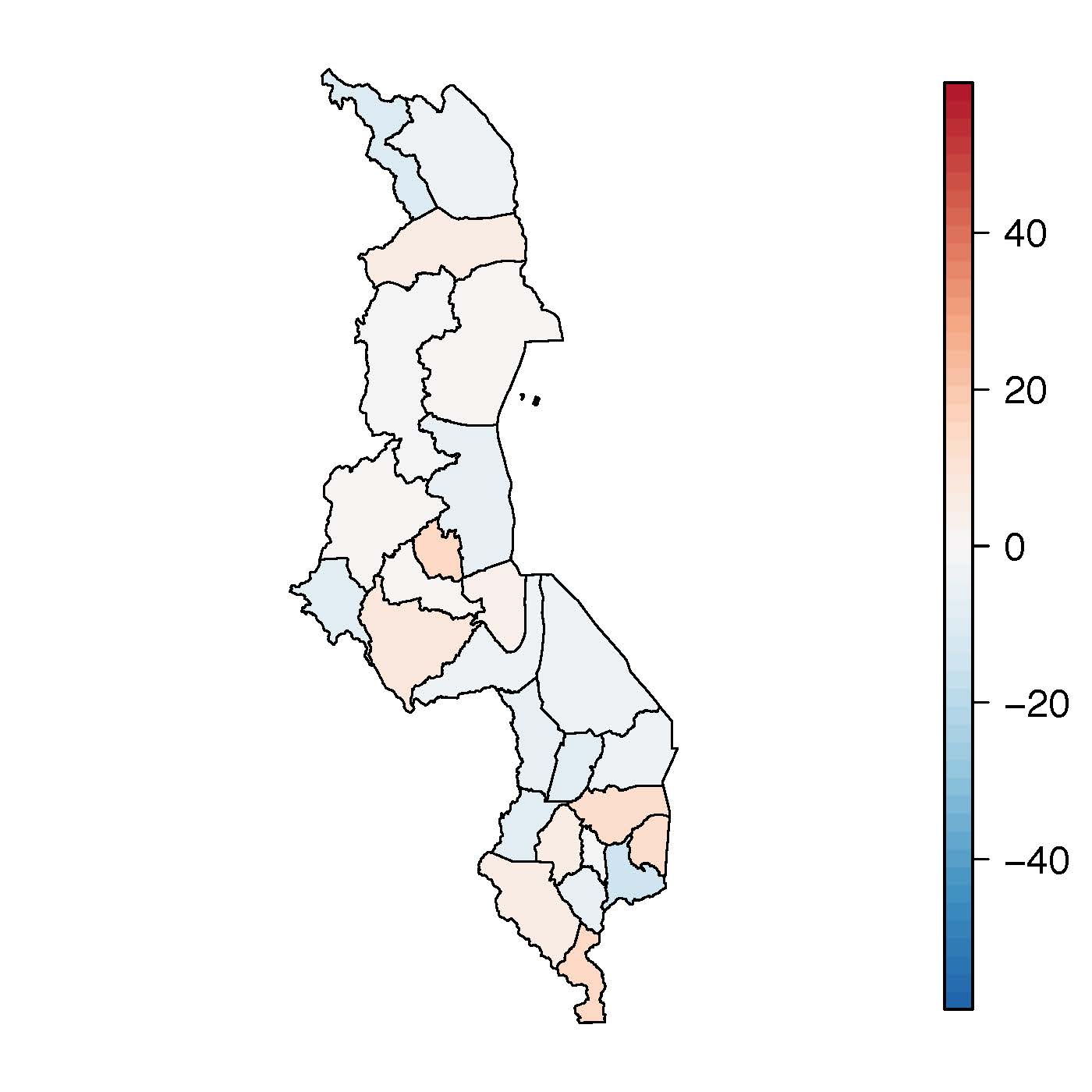}
  \vspace{-3mm}
 \subcaption{Education}\label{fig:edu}
  \end{subfigure}
    \begin{subfigure}[h]{\textwidth}
  \centering
  \includegraphics[width=.32\textwidth]{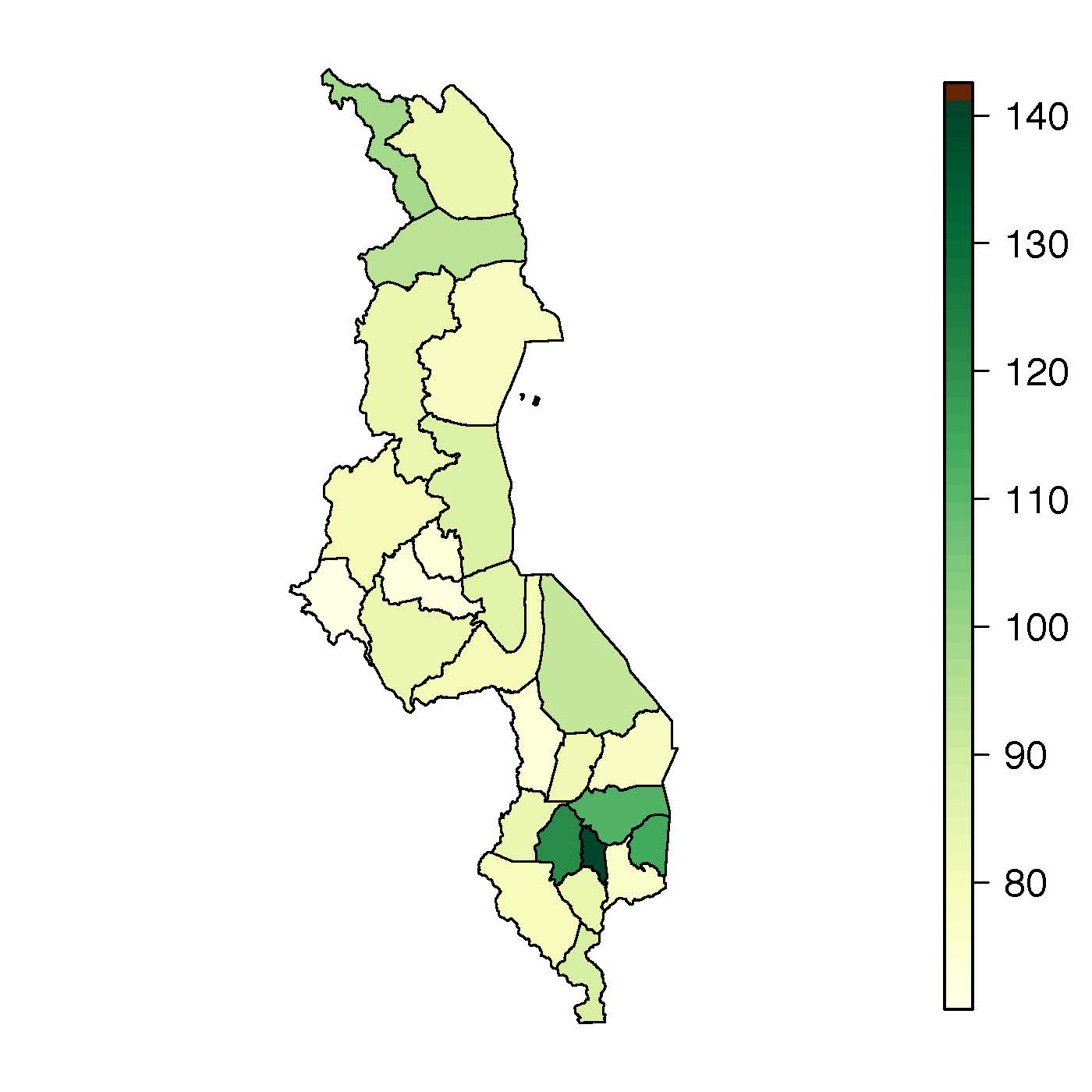}
  \includegraphics[width=.32\textwidth]{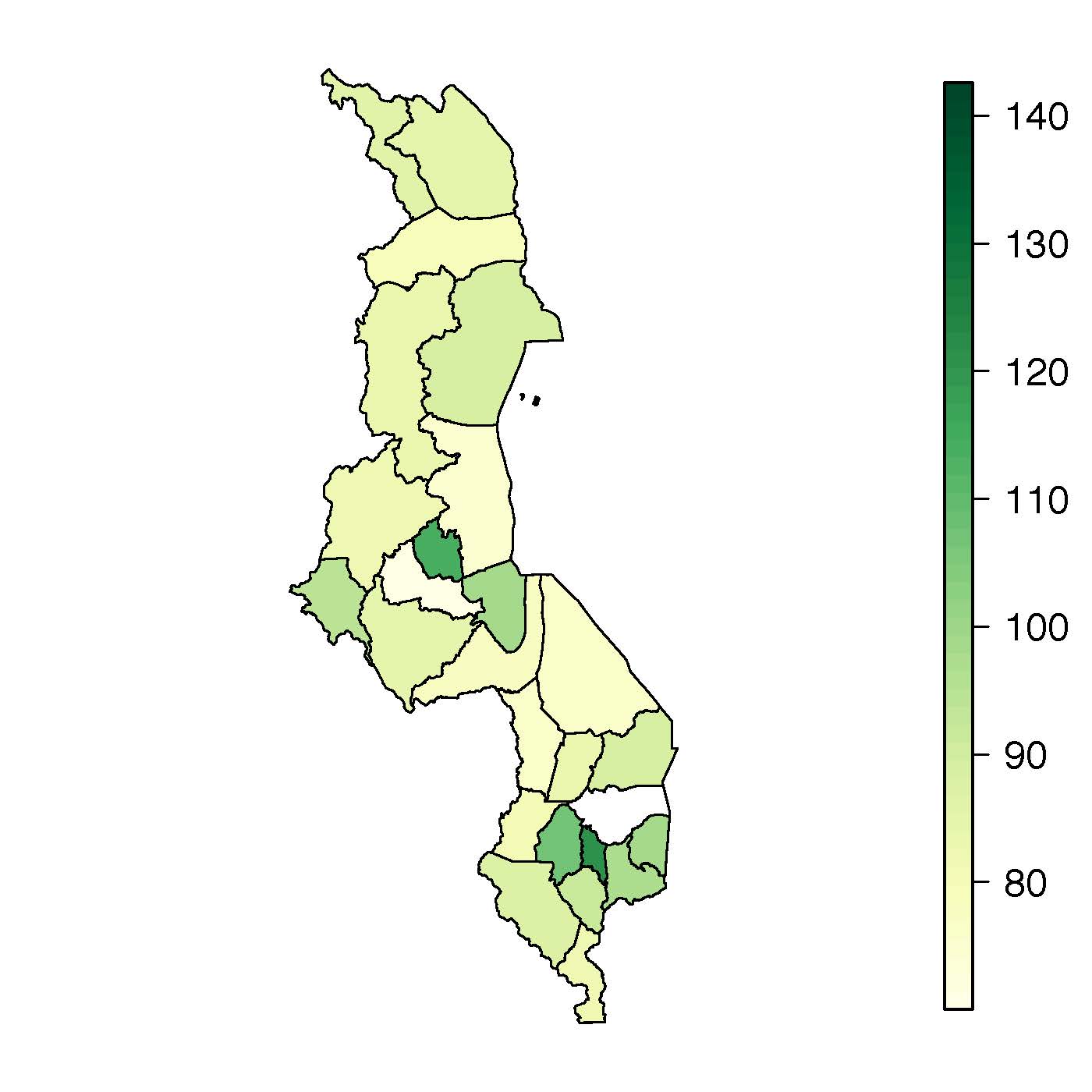}
  \includegraphics[width=.32\textwidth]{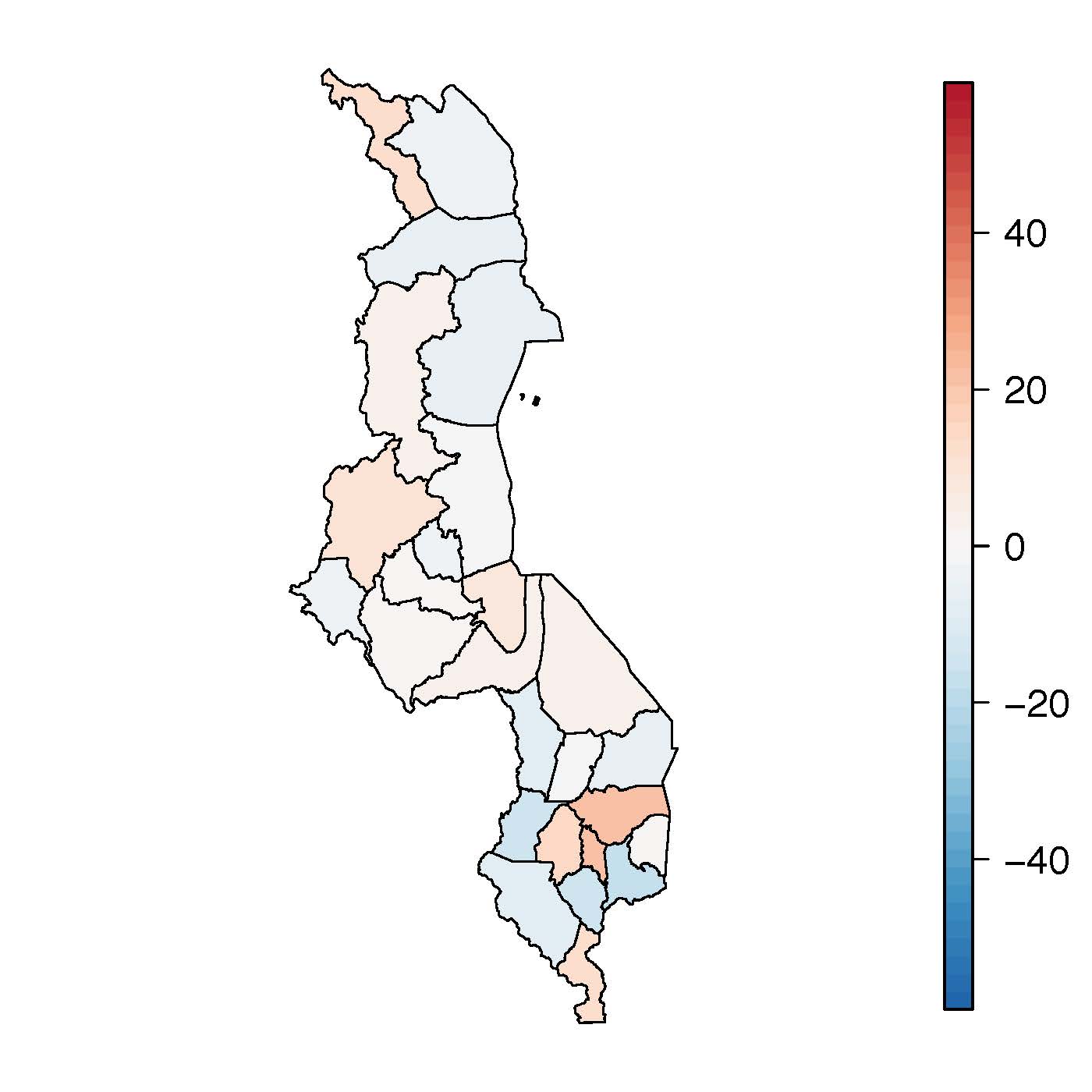}
  \vspace{-3mm}
 \subcaption{Health}\label{fig:health}
  \end{subfigure}
  \caption{Observed (left), predicted (center), and residual (right) donations plots for Agriculture, Education, and Health sectors. All values are in USD.}\label{fig:donplots2}
 \end{center}
\end{figure}

\begin{figure}[H]
 \begin{center}
    \begin{subfigure}[h]{\textwidth}
  \centering  
  \includegraphics[width=.32\textwidth]{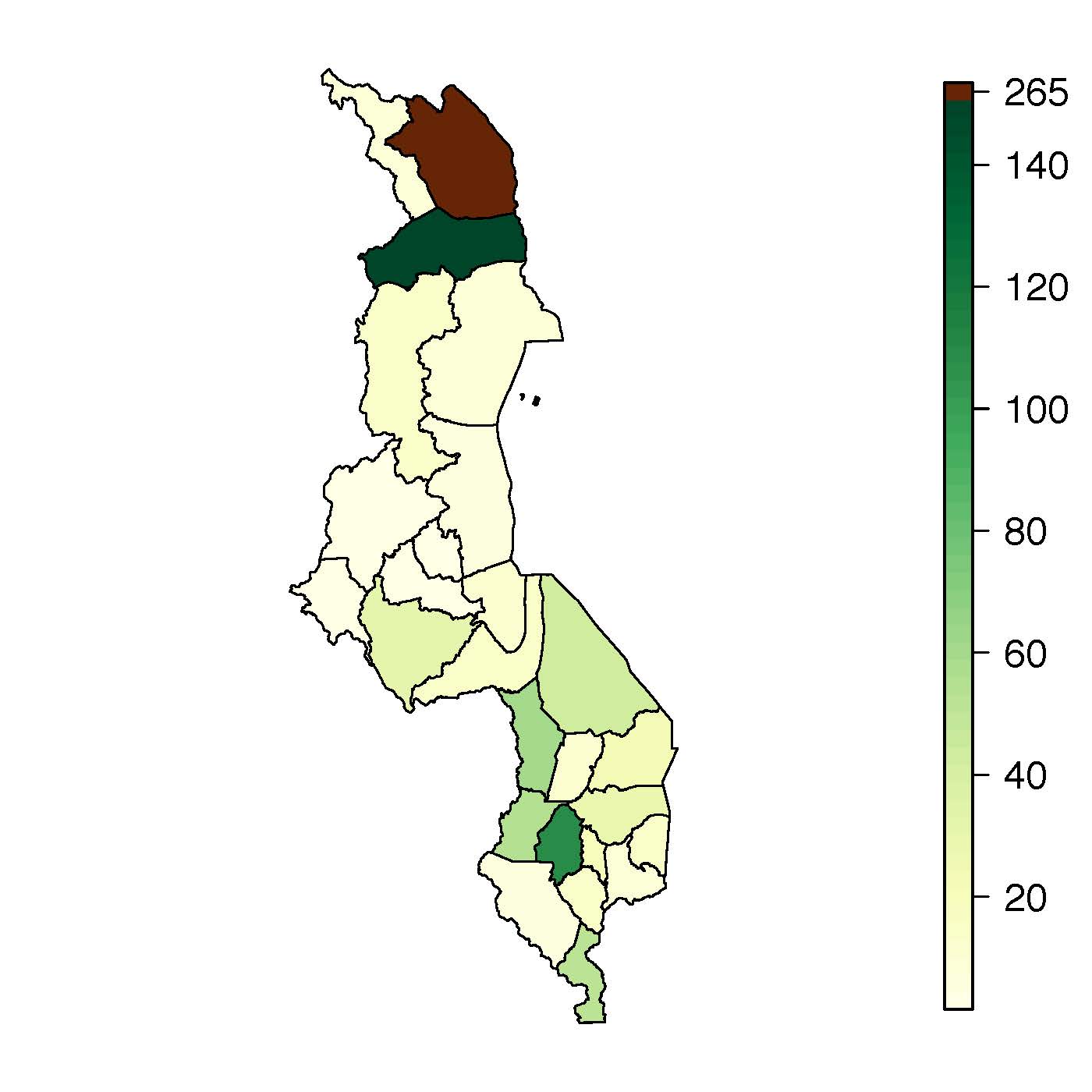}
  \includegraphics[width=.32\textwidth]{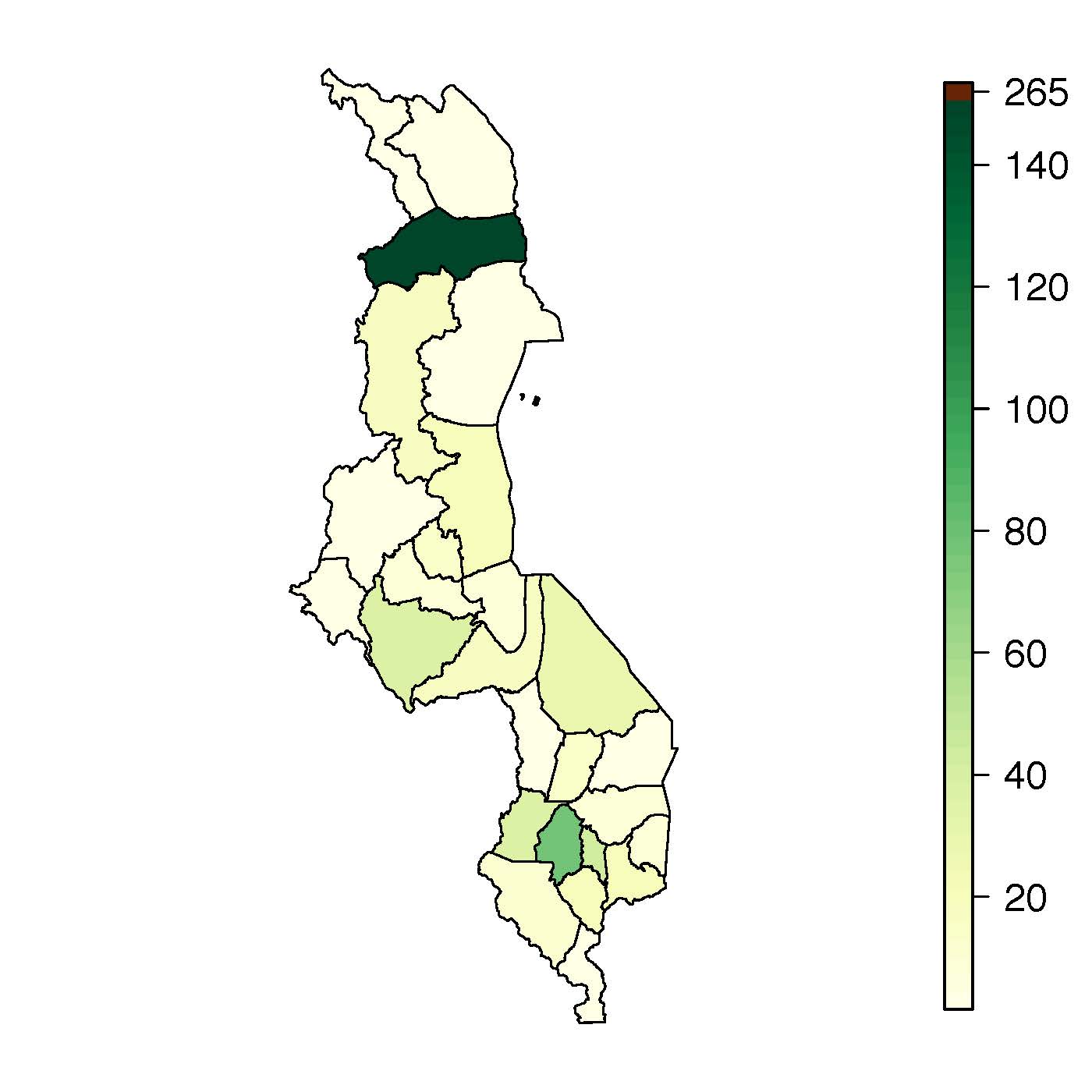}
  \includegraphics[width=.32\textwidth]{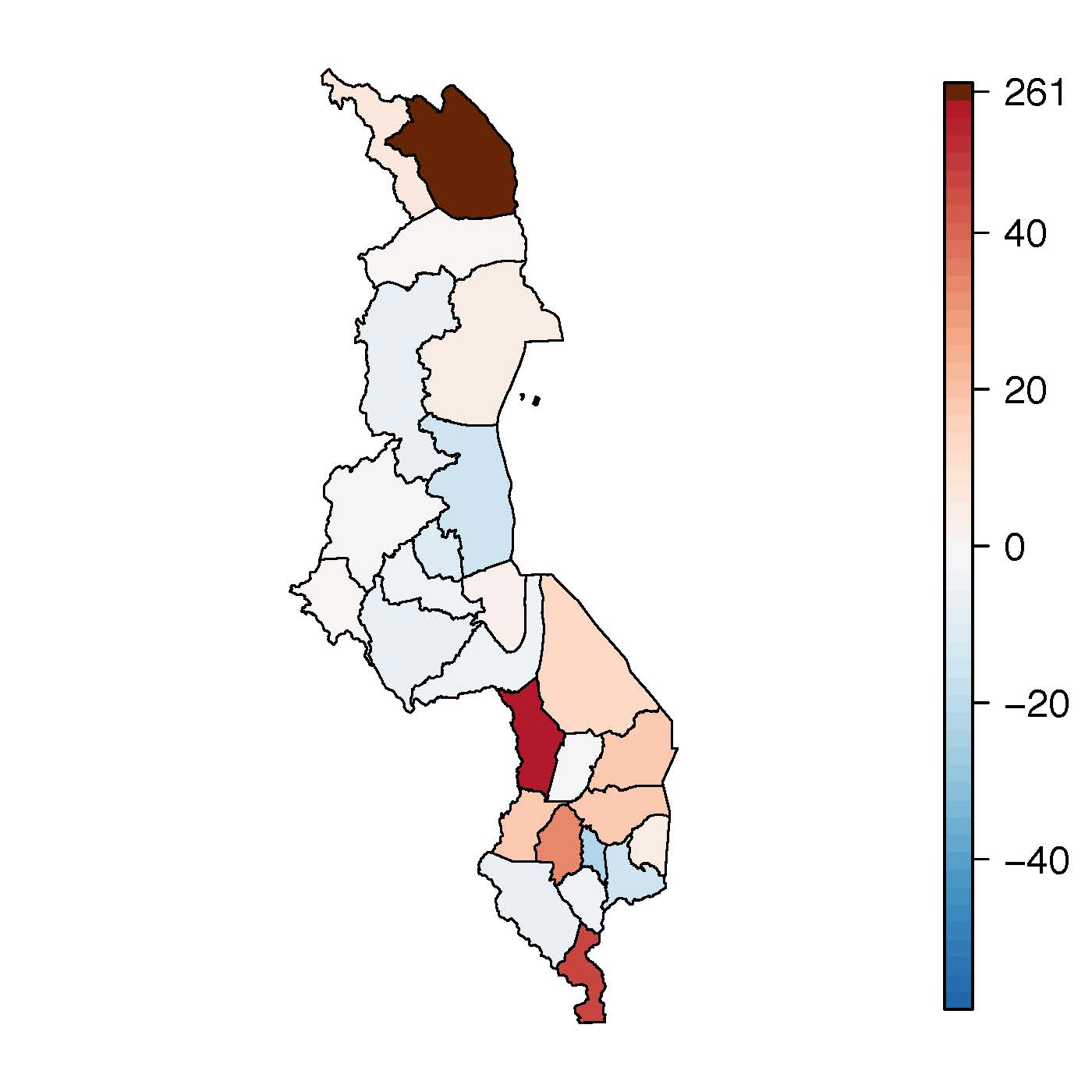}
  \vspace{-3mm}
 \subcaption{Roads, Public Works and Transportation}\label{fig:road}
 \end{subfigure}
   \begin{subfigure}[h]{\textwidth}
  \centering
    \includegraphics[width=.32\textwidth]{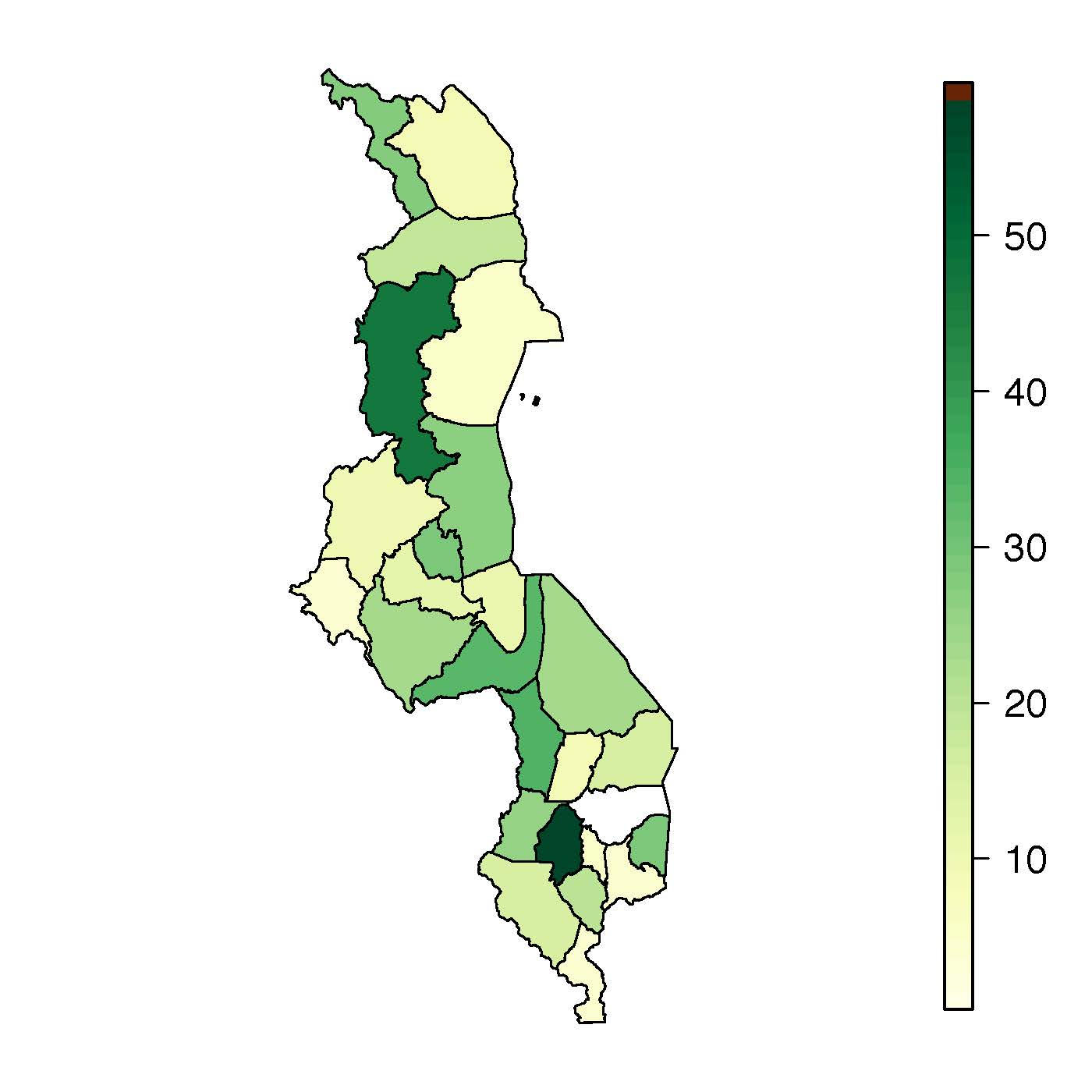}
  \includegraphics[width=.32\textwidth]{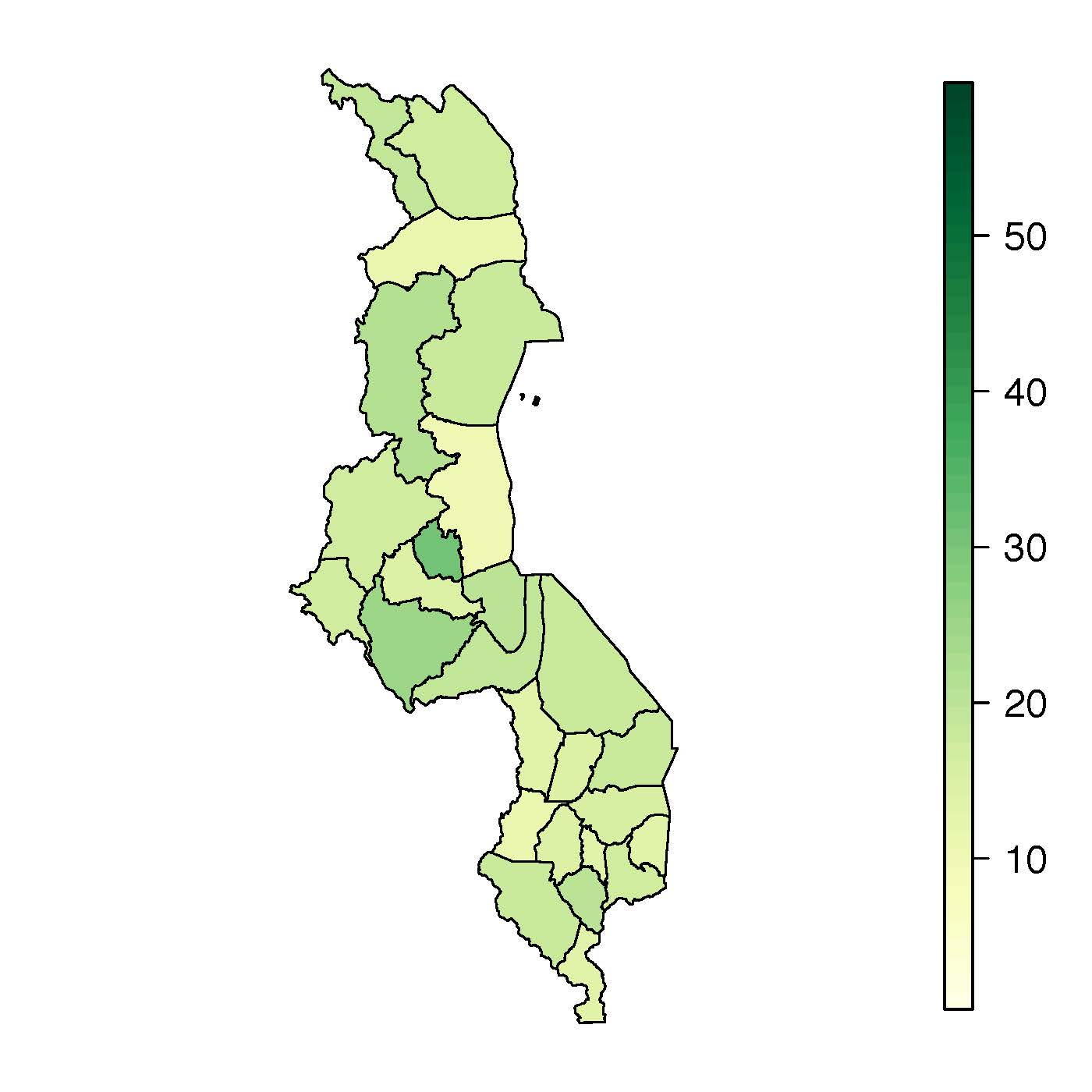}
  \includegraphics[width=.32\textwidth]{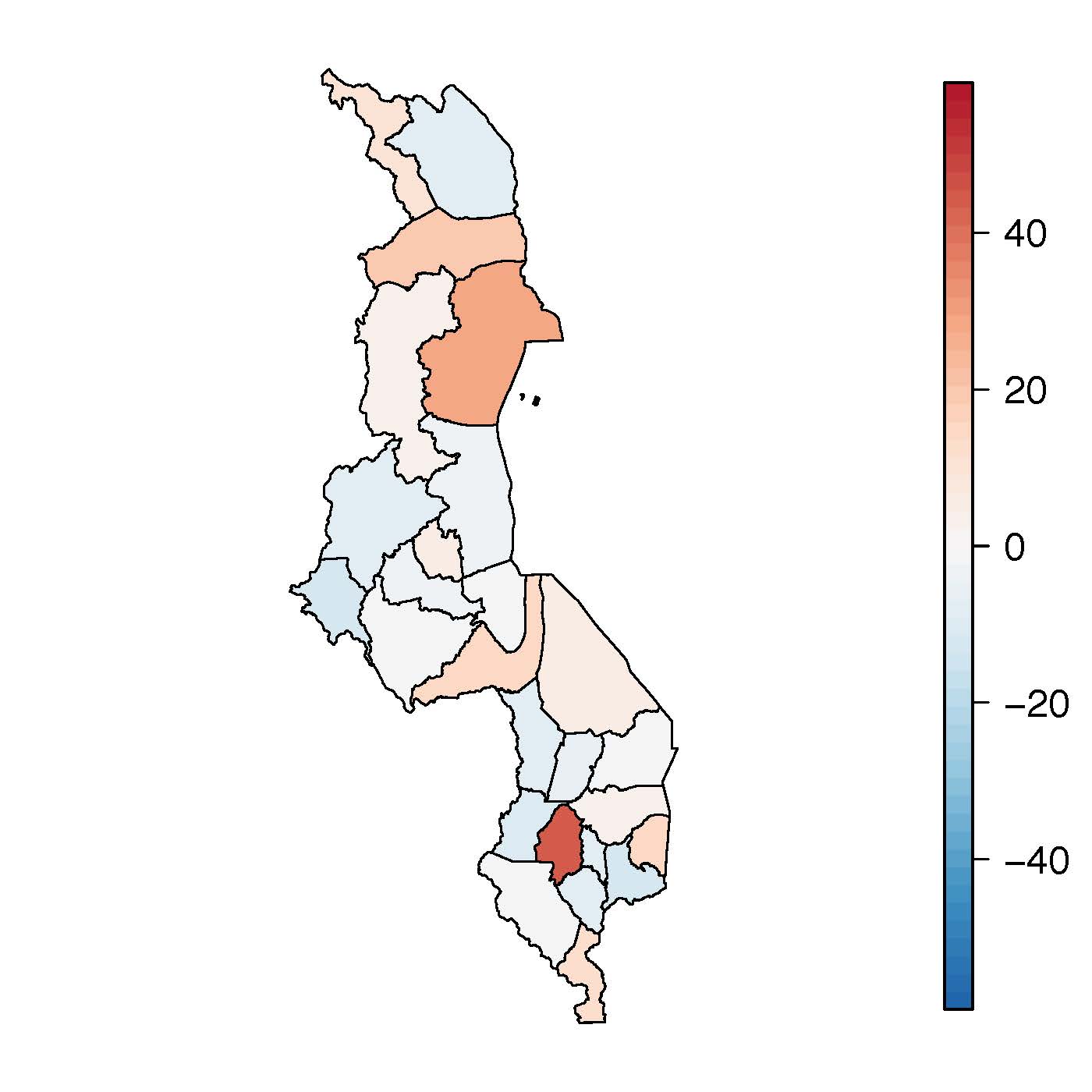}
  \vspace{-3mm}
 \caption{Water, Sanitation, and Irrigation}\label{fig:water}
  \end{subfigure}

  \caption{Observed (left), predicted (center), and residual (right) donations plots for RPT and WSI sectors. Outliers (in observed and residual donations of the RPT sector) are colored brown with values marked on the color legends.  All values are in USD.}\label{fig:donplots3}
 \end{center}
\end{figure}

Figures \ref{fig:donplots2} and \ref{fig:donplots3} provide additional plots of the observed, predicted, and residual values of donations per person for each sector.  

  \section{Coefficient Estimates}\label{app:coef}
  Tables \ref{tab:Agriculturecoeff}--\ref{tab:WSIcoeff} provide the posterior mean of each coefficient along with the corresponding monte carlo standard errors and posterior probabilities that the coefficient is greater than 0 for all models and sectors.
  \begin{landscape}
\begin{table} 
 \small \centering 
 \caption{Coefficient posterior means (monte carlo standard errors) for all models for the \emph{Agriculture} sector.  P(>0) means the posterior probability that the corresponding coefficient is greater than 0.  Coefficeints (other than intercepts) that have >90\% posterior probability of being either positive or negative are bolded.}\label{tab:Agriculturecoeff}
\begin{tabular}{c|cccccccc}\multicolumn{1}{r}{Covariate} & Intercept & Poverty & Dist to School & All Injured & Food \% & Land Cultivated & Electric \% & Population \\\hline
& mean (mcse) & mean (mcse) & mean (mcse) & mean (mcse) & mean (mcse) & mean (mcse) & mean (mcse) & mean (mcse) \\
Model & P(>0) & P(>0) & P(>0) & P(>0) & P(>0) & P(>0) & P(>0) & P(>0) \\ \hline \hline 
1  & 3.58 (0.001)&\textbf{ -0.1 } (0.001)&0.05 (0.002)&\textbf{ -0.14 } (0.001)&0.06 (0.001)&\textbf{ 0.11 } (0.001)&\textbf{ 0.19 } (0.001)&-0.05 (0.001)\\[-.03in]
&{\footnotesize 1 }&{\footnotesize \textbf{ 0.07 }}&{\footnotesize 0.7 }&{\footnotesize \textbf{ 0.02 }}&{\footnotesize 0.82 }&{\footnotesize \textbf{ 0.98 }}&{\footnotesize \textbf{ 1 }}&{\footnotesize 0.2 }\\
2  & 3.55 (0.002)&-0.08 (0.003)&0 (0.003)&\textbf{ -0.17 } (0.003)&0.1 (0.003)&\textbf{ 0.13 } (0.002)&\textbf{ 0.2 } (0.002)&-0.05 (0.002)\\[-.03in]
&{\footnotesize 1 }&{\footnotesize 0.22 }&{\footnotesize 0.49 }&{\footnotesize \textbf{ 0.06 }}&{\footnotesize 0.81 }&{\footnotesize \textbf{ 0.95 }}&{\footnotesize \textbf{ 0.99 }}&{\footnotesize 0.34 }\\
3  & 3.57 (0.001)&-0.12 (0.003)&0.05 (0.004)&-0.1 (0.003)&0.01 (0.002)&0.1 (0.002)&\textbf{ 0.18 } (0.002)&-0.05 (0.001)\\[-.03in]
&{\footnotesize 1 }&{\footnotesize 0.22 }&{\footnotesize 0.61 }&{\footnotesize 0.26 }&{\footnotesize 0.54 }&{\footnotesize 0.82 }&{\footnotesize \textbf{ 0.95 }}&{\footnotesize 0.34 }\\
4  & 3.57 (0.001)&-0.12 (0.004)&0.05 (0.005)&-0.09 (0.004)&0.01 (0.003)&0.11 (0.002)&0.18 (0.002)&-0.05 (0.002)\\[-.03in]
&{\footnotesize 1 }&{\footnotesize 0.26 }&{\footnotesize 0.58 }&{\footnotesize 0.31 }&{\footnotesize 0.54 }&{\footnotesize 0.77 }&{\footnotesize 0.9 }&{\footnotesize 0.35 }\\
5  & 37.77 (0.045)&-3.18 (0.141)&-0.51 (0.184)&-5.94 (0.132)&2.58 (0.104)&\textbf{ 6.64 } (0.074)&\textbf{ 10.24 } (0.067)&-1.9 (0.054)\\[-.03in]
&{\footnotesize 1 }&{\footnotesize 0.31 }&{\footnotesize 0.47 }&{\footnotesize 0.17 }&{\footnotesize 0.68 }&{\footnotesize \textbf{ 0.92 }}&{\footnotesize \textbf{ 0.99 }}&{\footnotesize 0.33 }\\
6  & 37.66 (0.058)&-3.39 (0.176)&-0.19 (0.221)&-5.99 (0.157)&2.44 (0.128)&6.64 (0.092)&\textbf{ 10.25 } (0.078)&-1.9 (0.071)\\[-.03in]
&{\footnotesize 1 }&{\footnotesize 0.34 }&{\footnotesize 0.49 }&{\footnotesize 0.22 }&{\footnotesize 0.64 }&{\footnotesize 0.86 }&{\footnotesize \textbf{ 0.97 }}&{\footnotesize 0.36 }\\
7  & 3.72 (0.003)&-0.12 (0.006)&0.03 (0.008)&-0.07 (0.006)&-0.01 (0.006)&0.14 (0.005)&\textbf{ 0.2 } (0.004)&-0.07 (0.002)\\[-.03in]
&{\footnotesize 1 }&{\footnotesize 0.2 }&{\footnotesize 0.58 }&{\footnotesize 0.3 }&{\footnotesize 0.47 }&{\footnotesize 0.86 }&{\footnotesize \textbf{ 0.98 }}&{\footnotesize 0.21 }\\
8  & 3.7 (0.001)&-0.11 (0.004)&0.01 (0.005)&-0.11 (0.004)&0 (0.003)&0.15 (0.002)&\textbf{ 0.21 } (0.002)&-0.05 (0.001)\\[-.03in]
&{\footnotesize 1 }&{\footnotesize 0.25 }&{\footnotesize 0.53 }&{\footnotesize 0.25 }&{\footnotesize 0.52 }&{\footnotesize 0.89 }&{\footnotesize \textbf{ 0.97 }}&{\footnotesize 0.31 }\\
\hline \end{tabular}\end{table}\begin{table} 
 \small \centering 
 \caption{Coefficient posterior means (monte carlo standard errors) for all models for the \emph{Education} sector.  P(>0) means the posterior probability that the corresponding coefficient is greater than 0.  Coefficeints (other than intercepts) that have >90\% posterior probability of being either positive or negative are bolded.}\label{tab:Educationcoeff}
\begin{tabular}{c|cccccccc}\multicolumn{1}{r}{Covariate} & Intercept & Poverty & Dist to School & All Injured & Food \% & Land Cultivated & Electric \% & Population \\\hline
& mean (mcse) & mean (mcse) & mean (mcse) & mean (mcse) & mean (mcse) & mean (mcse) & mean (mcse) & mean (mcse) \\
Model & P(>0) & P(>0) & P(>0) & P(>0) & P(>0) & P(>0) & P(>0) & P(>0) \\ \hline \hline 
1  & 3.51 (0.001)&\textbf{ 0.14 } (0.002)&\textbf{ -0.17 } (0.002)&\textbf{ 0.18 } (0.001)&-0.04 (0.002)&0.01 (0.001)&-0.04 (0.001)&-0.06 (0.001)\\[-.03in]
&{\footnotesize 1 }&{\footnotesize \textbf{ 0.96 }}&{\footnotesize \textbf{ 0.03 }}&{\footnotesize \textbf{ 0.99 }}&{\footnotesize 0.28 }&{\footnotesize 0.6 }&{\footnotesize 0.29 }&{\footnotesize 0.19 }\\
2  & 3.5 (0.001)&0.09 (0.003)&-0.16 (0.004)&\textbf{ 0.17 } (0.002)&0.04 (0.004)&-0.02 (0.001)&-0.03 (0.001)&-0.06 (0.001)\\[-.03in]
&{\footnotesize 1 }&{\footnotesize 0.8 }&{\footnotesize 0.13 }&{\footnotesize \textbf{ 0.97 }}&{\footnotesize 0.5 }&{\footnotesize 0.43 }&{\footnotesize 0.37 }&{\footnotesize 0.28 }\\
3  & 3.5 (0.001)&0.18 (0.004)&-0.2 (0.005)&\textbf{ 0.23 } (0.003)&-0.09 (0.002)&0 (0.002)&-0.02 (0.002)&-0.05 (0.001)\\[-.03in]
&{\footnotesize 1 }&{\footnotesize 0.87 }&{\footnotesize 0.14 }&{\footnotesize \textbf{ 0.93 }}&{\footnotesize 0.26 }&{\footnotesize 0.49 }&{\footnotesize 0.44 }&{\footnotesize 0.33 }\\
4  & 3.5 (0.001)&0.19 (0.004)&-0.2 (0.005)&0.23 (0.004)&-0.09 (0.003)&0 (0.002)&-0.02 (0.002)&-0.04 (0.002)\\[-.03in]
&{\footnotesize 1 }&{\footnotesize 0.83 }&{\footnotesize 0.18 }&{\footnotesize 0.89 }&{\footnotesize 0.29 }&{\footnotesize 0.5 }&{\footnotesize 0.45 }&{\footnotesize 0.36 }\\
5  & 34.61 (0.046)&6.44 (0.139)&-7.37 (0.184)&7.21 (0.129)&-2.61 (0.105)&0.16 (0.076)&-1.09 (0.066)&-2.22 (0.052)\\[-.03in]
&{\footnotesize 1 }&{\footnotesize 0.85 }&{\footnotesize 0.17 }&{\footnotesize 0.87 }&{\footnotesize 0.32 }&{\footnotesize 0.51 }&{\footnotesize 0.4 }&{\footnotesize 0.3 }\\
6  & 34.46 (0.059)&6.07 (0.173)&-7.07 (0.227)&7.22 (0.163)&-2.59 (0.126)&0.07 (0.1)&-1.1 (0.08)&-2.25 (0.068)\\[-.03in]
&{\footnotesize 1 }&{\footnotesize 0.78 }&{\footnotesize 0.23 }&{\footnotesize 0.82 }&{\footnotesize 0.35 }&{\footnotesize 0.5 }&{\footnotesize 0.42 }&{\footnotesize 0.34 }\\
7  & 3.61 (0.003)&\textbf{ 0.22 } (0.007)&\textbf{ -0.25 } (0.01)&\textbf{ 0.22 } (0.007)&-0.11 (0.005)&0.01 (0.004)&0 (0.004)&-0.05 (0.003)\\[-.03in]
&{\footnotesize 1 }&{\footnotesize \textbf{ 0.93 }}&{\footnotesize \textbf{ 0.1 }}&{\footnotesize \textbf{ 0.93 }}&{\footnotesize 0.19 }&{\footnotesize 0.55 }&{\footnotesize 0.46 }&{\footnotesize 0.28 }\\
8  & 3.63 (0.001)&0.21 (0.003)&\textbf{ -0.27 } (0.004)&\textbf{ 0.26 } (0.004)&-0.14 (0.003)&0 (0.002)&-0.01 (0.002)&-0.05 (0.001)\\[-.03in]
&{\footnotesize 1 }&{\footnotesize 0.9 }&{\footnotesize \textbf{ 0.09 }}&{\footnotesize \textbf{ 0.94 }}&{\footnotesize 0.16 }&{\footnotesize 0.52 }&{\footnotesize 0.45 }&{\footnotesize 0.3 }\\
\hline \end{tabular}\end{table}\begin{table} 
 \small \centering 
 \caption{Coefficient posterior means (monte carlo standard errors) for all models for the \emph{Governance} sector.  P(>0) means the posterior probability that the corresponding coefficient is greater than 0.  Coefficeints (other than intercepts) that have >90\% posterior probability of being either positive or negative are bolded.}\label{tab:Governancecoeff}
\begin{tabular}{c|cccccccc}\multicolumn{1}{r}{Covariate} & Intercept & Poverty & Dist to School & All Injured & Food \% & Land Cultivated & Electric \% & Population \\\hline
& mean (mcse) & mean (mcse) & mean (mcse) & mean (mcse) & mean (mcse) & mean (mcse) & mean (mcse) & mean (mcse) \\
Model & P(>0) & P(>0) & P(>0) & P(>0) & P(>0) & P(>0) & P(>0) & P(>0) \\ \hline \hline 
1  & 4.21 (0)&-0.02 (0.001)&0.03 (0.001)&0 (0.001)&-0.01 (0.001)&-0.02 (0)&0.03 (0)&\textbf{ 0.04 } (0)\\[-.03in]
&{\footnotesize 1 }&{\footnotesize 0.34 }&{\footnotesize 0.7 }&{\footnotesize 0.5 }&{\footnotesize 0.4 }&{\footnotesize 0.28 }&{\footnotesize 0.87 }&{\footnotesize \textbf{ 0.92 }}\\
2  & 4.21 (0)&-0.02 (0.001)&0.03 (0.002)&0 (0.001)&-0.01 (0.001)&-0.02 (0.001)&0.03 (0.001)&0.04 (0.001)\\[-.03in]
&{\footnotesize 1 }&{\footnotesize 0.37 }&{\footnotesize 0.66 }&{\footnotesize 0.52 }&{\footnotesize 0.42 }&{\footnotesize 0.31 }&{\footnotesize 0.8 }&{\footnotesize 0.85 }\\
3  & 4.21 (0.001)&-0.01 (0.003)&0.01 (0.004)&0 (0.003)&-0.01 (0.003)&-0.02 (0.002)&0.04 (0.002)&0.04 (0.001)\\[-.03in]
&{\footnotesize 1 }&{\footnotesize 0.48 }&{\footnotesize 0.53 }&{\footnotesize 0.51 }&{\footnotesize 0.46 }&{\footnotesize 0.44 }&{\footnotesize 0.63 }&{\footnotesize 0.64 }\\
4  & 4.21 (0.001)&-0.01 (0.004)&0.02 (0.005)&0 (0.004)&-0.02 (0.003)&-0.02 (0.002)&0.03 (0.002)&0.03 (0.002)\\[-.03in]
&{\footnotesize 1 }&{\footnotesize 0.48 }&{\footnotesize 0.53 }&{\footnotesize 0.5 }&{\footnotesize 0.47 }&{\footnotesize 0.45 }&{\footnotesize 0.6 }&{\footnotesize 0.6 }\\
5  & 67.6 (0.047)&-1.21 (0.142)&1.62 (0.182)&0.13 (0.126)&-1.02 (0.105)&-1.25 (0.077)&2.36 (0.064)&2.83 (0.053)\\[-.03in]
&{\footnotesize 1 }&{\footnotesize 0.43 }&{\footnotesize 0.58 }&{\footnotesize 0.51 }&{\footnotesize 0.43 }&{\footnotesize 0.4 }&{\footnotesize 0.7 }&{\footnotesize 0.74 }\\
6  & 67.58 (0.062)&-1.06 (0.174)&1.56 (0.231)&0.19 (0.16)&-1.17 (0.131)&-1.2 (0.098)&2.38 (0.08)&2.87 (0.067)\\[-.03in]
&{\footnotesize 1 }&{\footnotesize 0.44 }&{\footnotesize 0.56 }&{\footnotesize 0.52 }&{\footnotesize 0.43 }&{\footnotesize 0.42 }&{\footnotesize 0.67 }&{\footnotesize 0.7 }\\
7  & 4.54 (0.003)&0.04 (0.004)&-0.11 (0.007)&0.04 (0.005)&0.05 (0.006)&0.11 (0.006)&-0.06 (0.003)&0 (0.002)\\[-.03in]
&{\footnotesize 1 }&{\footnotesize 0.65 }&{\footnotesize 0.22 }&{\footnotesize 0.66 }&{\footnotesize 0.67 }&{\footnotesize 0.78 }&{\footnotesize 0.26 }&{\footnotesize 0.48 }\\
8  & 4.3 (0.001)&0 (0.003)&0 (0.004)&0 (0.003)&-0.05 (0.003)&-0.01 (0.002)&0.05 (0.002)&0.04 (0.001)\\[-.03in]
&{\footnotesize 1 }&{\footnotesize 0.5 }&{\footnotesize 0.49 }&{\footnotesize 0.5 }&{\footnotesize 0.38 }&{\footnotesize 0.48 }&{\footnotesize 0.65 }&{\footnotesize 0.63 }\\
\hline \end{tabular}\end{table}\begin{table} 
 \small \centering 
 \caption{Coefficient posterior means (monte carlo standard errors) for all models for the \emph{Health} sector.  P(>0) means the posterior probability that the corresponding coefficient is greater than 0.  Coefficeints (other than intercepts) that have >90\% posterior probability of being either positive or negative are bolded.}\label{tab:Healthcoeff}
\begin{tabular}{c|cccccccc}\multicolumn{1}{r}{Covariate} & Intercept & Poverty & Dist to School & All Injured & Food \% & Land Cultivated & Electric \% & Population \\\hline
& mean (mcse) & mean (mcse) & mean (mcse) & mean (mcse) & mean (mcse) & mean (mcse) & mean (mcse) & mean (mcse) \\
Model & P(>0) & P(>0) & P(>0) & P(>0) & P(>0) & P(>0) & P(>0) & P(>0) \\ \hline \hline 
1  & 4.47 (0)&\textbf{ 0.07 } (0.001)&-0.02 (0.001)&0.01 (0.001)&0 (0.001)&\textbf{ -0.11 } (0)&\textbf{ 0.04 } (0)&-0.02 (0)\\[-.03in]
&{\footnotesize 1 }&{\footnotesize \textbf{ 0.99 }}&{\footnotesize 0.31 }&{\footnotesize 0.6 }&{\footnotesize 0.46 }&{\footnotesize \textbf{ 0 }}&{\footnotesize \textbf{ 0.97 }}&{\footnotesize 0.18 }\\
2  & 4.46 (0)&\textbf{ 0.07 } (0.001)&-0.02 (0.001)&0.01 (0.001)&0 (0.001)&\textbf{ -0.11 } (0.001)&\textbf{ 0.04 } (0)&-0.02 (0)\\[-.03in]
&{\footnotesize 1 }&{\footnotesize \textbf{ 0.97 }}&{\footnotesize 0.31 }&{\footnotesize 0.65 }&{\footnotesize 0.43 }&{\footnotesize \textbf{ 0 }}&{\footnotesize \textbf{ 0.93 }}&{\footnotesize 0.25 }\\
3  & 4.47 (0.001)&0.07 (0.003)&-0.03 (0.005)&0.02 (0.003)&0 (0.003)&-0.1 (0.002)&0.04 (0.002)&-0.02 (0.001)\\[-.03in]
&{\footnotesize 1 }&{\footnotesize 0.66 }&{\footnotesize 0.44 }&{\footnotesize 0.55 }&{\footnotesize 0.5 }&{\footnotesize 0.2 }&{\footnotesize 0.64 }&{\footnotesize 0.43 }\\
4  & 4.47 (0.001)&0.07 (0.004)&-0.04 (0.006)&0.03 (0.004)&-0.01 (0.003)&-0.09 (0.002)&0.04 (0.002)&-0.02 (0.002)\\[-.03in]
&{\footnotesize 1 }&{\footnotesize 0.65 }&{\footnotesize 0.43 }&{\footnotesize 0.56 }&{\footnotesize 0.48 }&{\footnotesize 0.26 }&{\footnotesize 0.63 }&{\footnotesize 0.43 }\\
5  & 88.67 (0.048)&7.43 (0.136)&-3.23 (0.173)&2.02 (0.125)&-0.65 (0.104)&\textbf{ -9.57 } (0.074)&3.69 (0.063)&-2.1 (0.054)\\[-.03in]
&{\footnotesize 1 }&{\footnotesize 0.88 }&{\footnotesize 0.33 }&{\footnotesize 0.63 }&{\footnotesize 0.45 }&{\footnotesize \textbf{ 0.02 }}&{\footnotesize 0.79 }&{\footnotesize 0.31 }\\
6  & 88.62 (0.059)&7.42 (0.178)&-3.07 (0.232)&1.99 (0.155)&-0.66 (0.126)&\textbf{ -9.57 } (0.099)&3.59 (0.083)&-2.3 (0.071)\\[-.03in]
&{\footnotesize 1 }&{\footnotesize 0.82 }&{\footnotesize 0.37 }&{\footnotesize 0.61 }&{\footnotesize 0.47 }&{\footnotesize \textbf{ 0.06 }}&{\footnotesize 0.74 }&{\footnotesize 0.33 }\\
7  & 4.58 (0.003)&0.07 (0.008)&-0.04 (0.01)&0 (0.007)&-0.01 (0.006)&-0.08 (0.004)&0.05 (0.003)&-0.03 (0.003)\\[-.03in]
&{\footnotesize 1 }&{\footnotesize 0.65 }&{\footnotesize 0.44 }&{\footnotesize 0.48 }&{\footnotesize 0.49 }&{\footnotesize 0.24 }&{\footnotesize 0.67 }&{\footnotesize 0.34 }\\
8  & 4.57 (0.001)&0.08 (0.004)&-0.04 (0.005)&0.02 (0.003)&-0.04 (0.003)&-0.1 (0.002)&0.05 (0.002)&-0.03 (0.001)\\[-.03in]
&{\footnotesize 1 }&{\footnotesize 0.69 }&{\footnotesize 0.43 }&{\footnotesize 0.54 }&{\footnotesize 0.41 }&{\footnotesize 0.2 }&{\footnotesize 0.65 }&{\footnotesize 0.38 }\\
\hline \end{tabular}\end{table}\begin{table} 
 \small \centering 
 \caption{Coefficient posterior means (monte carlo standard errors) for all models for the \emph{RD} sector.  P(>0) means the posterior probability that the corresponding coefficient is greater than 0.  Coefficeints (other than intercepts) that have >90\% posterior probability of being either positive or negative are bolded.}\label{tab:RDcoeff}
\begin{tabular}{c|cccccccc}\multicolumn{1}{r}{Covariate} & Intercept & Poverty & Dist to School & All Injured & Food \% & Land Cultivated & Electric \% & Population \\\hline
& mean (mcse) & mean (mcse) & mean (mcse) & mean (mcse) & mean (mcse) & mean (mcse) & mean (mcse) & mean (mcse) \\
Model & P(>0) & P(>0) & P(>0) & P(>0) & P(>0) & P(>0) & P(>0) & P(>0) \\ \hline \hline 
1  & 2.6 (0.002)&\textbf{ -0.27 } (0.004)&\textbf{ 0.41 } (0.005)&-0.12 (0.002)&0.05 (0.004)&-0.07 (0.002)&0.09 (0.002)&-0.1 (0.002)\\[-.03in]
&{\footnotesize 1 }&{\footnotesize \textbf{ 0.06 }}&{\footnotesize \textbf{ 0.99 }}&{\footnotesize 0.14 }&{\footnotesize 0.58 }&{\footnotesize 0.25 }&{\footnotesize 0.82 }&{\footnotesize 0.22 }\\
2  & 2.69 (0.002)&\textbf{ -0.29 } (0.004)&\textbf{ 0.42 } (0.005)&-0.14 (0.003)&0.06 (0.003)&-0.08 (0.002)&0.08 (0.001)&-0.11 (0.001)\\[-.03in]
&{\footnotesize 1 }&{\footnotesize \textbf{ 0.04 }}&{\footnotesize \textbf{ 0.98 }}&{\footnotesize 0.17 }&{\footnotesize 0.64 }&{\footnotesize 0.23 }&{\footnotesize 0.81 }&{\footnotesize 0.13 }\\
3  & 2.36 (0.001)&\textbf{ -0.52 } (0.003)&\textbf{ 0.76 } (0.004)&\textbf{ -0.19 } (0.003)&-0.01 (0.003)&-0.07 (0.002)&\textbf{ 0.15 } (0.002)&\textbf{ -0.18 } (0.001)\\[-.03in]
&{\footnotesize 1 }&{\footnotesize \textbf{ 0 }}&{\footnotesize \textbf{ 1 }}&{\footnotesize \textbf{ 0.09 }}&{\footnotesize 0.48 }&{\footnotesize 0.28 }&{\footnotesize \textbf{ 0.92 }}&{\footnotesize \textbf{ 0.04 }}\\
4  & 2.36 (0.001)&\textbf{ -0.53 } (0.004)&\textbf{ 0.76 } (0.005)&-0.2 (0.004)&0 (0.003)&-0.06 (0.002)&0.15 (0.002)&\textbf{ -0.18 } (0.002)\\[-.03in]
&{\footnotesize 1 }&{\footnotesize \textbf{ 0 }}&{\footnotesize \textbf{ 1 }}&{\footnotesize 0.14 }&{\footnotesize 0.5 }&{\footnotesize 0.33 }&{\footnotesize 0.85 }&{\footnotesize \textbf{ 0.09 }}\\
5  & 13.33 (0.047)&-4.03 (0.141)&6.95 (0.179)&-2.52 (0.127)&0.67 (0.102)&-1.82 (0.073)&1.73 (0.061)&-2.32 (0.054)\\[-.03in]
&{\footnotesize 1 }&{\footnotesize 0.27 }&{\footnotesize 0.82 }&{\footnotesize 0.34 }&{\footnotesize 0.55 }&{\footnotesize 0.35 }&{\footnotesize 0.65 }&{\footnotesize 0.3 }\\
6  & 13.39 (0.058)&-4.27 (0.18)&7.32 (0.234)&-2.71 (0.157)&0.68 (0.123)&-1.85 (0.096)&1.73 (0.079)&-2.46 (0.068)\\[-.03in]
&{\footnotesize 1 }&{\footnotesize 0.3 }&{\footnotesize 0.77 }&{\footnotesize 0.36 }&{\footnotesize 0.54 }&{\footnotesize 0.38 }&{\footnotesize 0.62 }&{\footnotesize 0.32 }\\
7  & 2.6 (0.003)&\textbf{ -0.54 } (0.005)&\textbf{ 0.72 } (0.008)&-0.07 (0.006)&-0.04 (0.005)&-0.07 (0.004)&0.14 (0.004)&\textbf{ -0.19 } (0.003)\\[-.03in]
&{\footnotesize 1 }&{\footnotesize \textbf{ 0 }}&{\footnotesize \textbf{ 1 }}&{\footnotesize 0.31 }&{\footnotesize 0.39 }&{\footnotesize 0.27 }&{\footnotesize 0.9 }&{\footnotesize \textbf{ 0.04 }}\\
8  & 2.71 (0.002)&\textbf{ -0.29 } (0.006)&\textbf{ 0.56 } (0.007)&-0.21 (0.004)&0.03 (0.003)&-0.07 (0.003)&0.09 (0.002)&-0.12 (0.003)\\[-.03in]
&{\footnotesize 1 }&{\footnotesize \textbf{ 0.1 }}&{\footnotesize \textbf{ 0.98 }}&{\footnotesize 0.12 }&{\footnotesize 0.59 }&{\footnotesize 0.32 }&{\footnotesize 0.76 }&{\footnotesize 0.22 }\\
\hline \end{tabular}\end{table}\begin{table} 
 \small \centering 
 \caption{Coefficient posterior means (monte carlo standard errors) for all models for the \emph{RPT} sector.  P(>0) means the posterior probability that the corresponding coefficient is greater than 0.  Coefficeints (other than intercepts) that have >90\% posterior probability of being either positive or negative are bolded.}\label{tab:RPTcoeff}
\begin{tabular}{c|cccccccc}\multicolumn{1}{r}{Covariate} & Intercept & Poverty & Dist to School & All Injured & Food \% & Land Cultivated & Electric \% & Population \\\hline
& mean (mcse) & mean (mcse) & mean (mcse) & mean (mcse) & mean (mcse) & mean (mcse) & mean (mcse) & mean (mcse) \\
Model & P(>0) & P(>0) & P(>0) & P(>0) & P(>0) & P(>0) & P(>0) & P(>0) \\ \hline \hline 
1  & 2.11 (0.004)&\textbf{ 0.89 } (0.003)&-0.07 (0.006)&\textbf{ -0.39 } (0.004)&0.21 (0.004)&0.11 (0.003)&\textbf{ 0.7 } (0.002)&\textbf{ 0.43 } (0.002)\\[-.03in]
&{\footnotesize 1 }&{\footnotesize \textbf{ 1 }}&{\footnotesize 0.38 }&{\footnotesize \textbf{ 0.01 }}&{\footnotesize 0.9 }&{\footnotesize 0.8 }&{\footnotesize \textbf{ 1 }}&{\footnotesize \textbf{ 1 }}\\
2  & 2.54 (0.005)&\textbf{ 0.34 } (0.006)&-0.11 (0.006)&\textbf{ 0.59 } (0.004)&\textbf{ -0.68 } (0.005)&\textbf{ -0.52 } (0.003)&\textbf{ 0.94 } (0.004)&\textbf{ 0.39 } (0.003)\\[-.03in]
&{\footnotesize 1 }&{\footnotesize \textbf{ 0.94 }}&{\footnotesize 0.31 }&{\footnotesize \textbf{ 1 }}&{\footnotesize \textbf{ 0 }}&{\footnotesize \textbf{ 0 }}&{\footnotesize \textbf{ 1 }}&{\footnotesize \textbf{ 1 }}\\
3  & 2.98 (0.001)&-0.2 (0.003)&0.23 (0.004)&-0.17 (0.003)&\textbf{ 0.32 } (0.002)&\textbf{ -0.47 } (0.002)&\textbf{ 0.32 } (0.002)&-0.03 (0.001)\\[-.03in]
&{\footnotesize 1 }&{\footnotesize 0.1 }&{\footnotesize 0.89 }&{\footnotesize 0.12 }&{\footnotesize \textbf{ 0.99 }}&{\footnotesize \textbf{ 0 }}&{\footnotesize \textbf{ 1 }}&{\footnotesize 0.4 }\\
4  & 2.98 (0.001)&-0.19 (0.004)&0.22 (0.005)&-0.16 (0.004)&\textbf{ 0.31 } (0.003)&\textbf{ -0.47 } (0.002)&\textbf{ 0.32 } (0.002)&-0.02 (0.002)\\[-.03in]
&{\footnotesize 1 }&{\footnotesize 0.16 }&{\footnotesize 0.83 }&{\footnotesize 0.2 }&{\footnotesize \textbf{ 0.96 }}&{\footnotesize \textbf{ 0 }}&{\footnotesize \textbf{ 0.99 }}&{\footnotesize 0.44 }\\
5  & 38.63 (0.045)&\textbf{ -23.6 } (0.14)&\textbf{ 19.33 } (0.179)&-7.41 (0.128)&\textbf{ 10.82 } (0.104)&\textbf{ -11.87 } (0.073)&\textbf{ 13.93 } (0.064)&\textbf{ -8.57 } (0.055)\\[-.03in]
&{\footnotesize 1 }&{\footnotesize \textbf{ 0 }}&{\footnotesize \textbf{ 1 }}&{\footnotesize 0.12 }&{\footnotesize \textbf{ 0.97 }}&{\footnotesize \textbf{ 0.01 }}&{\footnotesize \textbf{ 1 }}&{\footnotesize \textbf{ 0.02 }}\\
6  & 38.56 (0.057)&\textbf{ -23.36 } (0.176)&\textbf{ 18.92 } (0.222)&-7.48 (0.161)&\textbf{ 10.92 } (0.127)&\textbf{ -11.68 } (0.094)&\textbf{ 13.96 } (0.081)&\textbf{ -8.69 } (0.072)\\[-.03in]
&{\footnotesize 1 }&{\footnotesize \textbf{ 0 }}&{\footnotesize \textbf{ 0.98 }}&{\footnotesize 0.17 }&{\footnotesize \textbf{ 0.94 }}&{\footnotesize \textbf{ 0.03 }}&{\footnotesize \textbf{ 0.99 }}&{\footnotesize \textbf{ 0.06 }}\\
7  & 3.48 (0.003)&\textbf{ -0.44 } (0.006)&\textbf{ 0.47 } (0.009)&\textbf{ -0.25 } (0.006)&\textbf{ 0.48 } (0.005)&\textbf{ -0.52 } (0.004)&\textbf{ 0.19 } (0.004)&\textbf{ -0.21 } (0.003)\\[-.03in]
&{\footnotesize 1 }&{\footnotesize \textbf{ 0 }}&{\footnotesize \textbf{ 0.98 }}&{\footnotesize \textbf{ 0.03 }}&{\footnotesize \textbf{ 1 }}&{\footnotesize \textbf{ 0 }}&{\footnotesize \textbf{ 0.94 }}&{\footnotesize \textbf{ 0.03 }}\\
8  & 3.71 (0.002)&\textbf{ -0.66 } (0.004)&\textbf{ 0.48 } (0.007)&\textbf{ -0.62 } (0.004)&\textbf{ 0.71 } (0.003)&\textbf{ -0.33 } (0.004)&\textbf{ 0.28 } (0.003)&\textbf{ -0.18 } (0.002)\\[-.03in]
&{\footnotesize 1 }&{\footnotesize \textbf{ 0 }}&{\footnotesize \textbf{ 0.93 }}&{\footnotesize \textbf{ 0 }}&{\footnotesize \textbf{ 1 }}&{\footnotesize \textbf{ 0.04 }}&{\footnotesize \textbf{ 0.99 }}&{\footnotesize \textbf{ 0.06 }}\\
\hline \end{tabular}\end{table}\begin{table} 
 \small \centering 
 \caption{Coefficient posterior means (monte carlo standard errors) for all models for the \emph{WSI} sector.  P(>0) means the posterior probability that the corresponding coefficient is greater than 0.  Coefficeints (other than intercepts) that have >90\% posterior probability of being either positive or negative are bolded.}\label{tab:WSIcoeff}
\begin{tabular}{c|cccccccc}\multicolumn{1}{r}{Covariate} & Intercept & Poverty & Dist to School & All Injured & Food \% & Land Cultivated & Electric \% & Population \\\hline
& mean (mcse) & mean (mcse) & mean (mcse) & mean (mcse) & mean (mcse) & mean (mcse) & mean (mcse) & mean (mcse) \\
Model & P(>0) & P(>0) & P(>0) & P(>0) & P(>0) & P(>0) & P(>0) & P(>0) \\ \hline \hline 
1  & 2.66 (0.002)&0.06 (0.003)&0 (0.004)&0.06 (0.004)&-0.1 (0.003)&\textbf{ 0.15 } (0.002)&-0.17 (0.003)&0.05 (0.002)\\[-.03in]
&{\footnotesize 1 }&{\footnotesize 0.65 }&{\footnotesize 0.51 }&{\footnotesize 0.64 }&{\footnotesize 0.2 }&{\footnotesize \textbf{ 0.9 }}&{\footnotesize 0.16 }&{\footnotesize 0.69 }\\
2  & 2.76 (0.002)&-0.07 (0.006)&0.21 (0.006)&0.06 (0.003)&-0.15 (0.004)&-0.03 (0.002)&-0.15 (0.004)&0.09 (0.002)\\[-.03in]
&{\footnotesize 1 }&{\footnotesize 0.37 }&{\footnotesize 0.81 }&{\footnotesize 0.66 }&{\footnotesize 0.16 }&{\footnotesize 0.38 }&{\footnotesize 0.19 }&{\footnotesize 0.78 }\\
3  & 2.67 (0.001)&0.03 (0.003)&0.13 (0.004)&\textbf{ 0.26 } (0.003)&\textbf{ -0.39 } (0.003)&-0.12 (0.002)&-0.12 (0.002)&0.08 (0.001)\\[-.03in]
&{\footnotesize 1 }&{\footnotesize 0.58 }&{\footnotesize 0.76 }&{\footnotesize \textbf{ 0.95 }}&{\footnotesize \textbf{ 0 }}&{\footnotesize 0.14 }&{\footnotesize 0.14 }&{\footnotesize 0.78 }\\
4  & 2.66 (0.001)&0.04 (0.004)&0.12 (0.006)&\textbf{ 0.26 } (0.004)&\textbf{ -0.39 } (0.003)&-0.12 (0.002)&-0.11 (0.002)&0.08 (0.002)\\[-.03in]
&{\footnotesize 1 }&{\footnotesize 0.58 }&{\footnotesize 0.7 }&{\footnotesize \textbf{ 0.91 }}&{\footnotesize \textbf{ 0.01 }}&{\footnotesize 0.21 }&{\footnotesize 0.21 }&{\footnotesize 0.73 }\\
5  & 19.77 (0.048)&1.04 (0.137)&3.49 (0.183)&4.16 (0.126)&-7.18 (0.104)&-3.31 (0.078)&0.18 (0.067)&-0.63 (0.055)\\[-.03in]
&{\footnotesize 1 }&{\footnotesize 0.57 }&{\footnotesize 0.67 }&{\footnotesize 0.75 }&{\footnotesize 0.1 }&{\footnotesize 0.25 }&{\footnotesize 0.51 }&{\footnotesize 0.44 }\\
6  & 19.75 (0.057)&1.01 (0.177)&3.5 (0.226)&4.26 (0.164)&-7.2 (0.129)&-3.47 (0.092)&0.11 (0.083)&-0.72 (0.068)\\[-.03in]
&{\footnotesize 1 }&{\footnotesize 0.55 }&{\footnotesize 0.63 }&{\footnotesize 0.71 }&{\footnotesize 0.15 }&{\footnotesize 0.29 }&{\footnotesize 0.5 }&{\footnotesize 0.45 }\\
7  & 3.07 (0.003)&-0.09 (0.007)&\textbf{ 0.3 } (0.009)&\textbf{ 0.32 } (0.007)&\textbf{ -0.31 } (0.006)&\textbf{ -0.32 } (0.005)&-0.01 (0.003)&-0.08 (0.003)\\[-.03in]
&{\footnotesize 1 }&{\footnotesize 0.26 }&{\footnotesize \textbf{ 0.95 }}&{\footnotesize \textbf{ 0.97 }}&{\footnotesize \textbf{ 0.01 }}&{\footnotesize \textbf{ 0 }}&{\footnotesize 0.44 }&{\footnotesize 0.19 }\\
8  & 3.16 (0.001)&0.04 (0.002)&0.16 (0.004)&\textbf{ 0.36 } (0.004)&\textbf{ -0.39 } (0.003)&\textbf{ -0.33 } (0.003)&-0.02 (0.001)&-0.07 (0.001)\\[-.03in]
&{\footnotesize 1 }&{\footnotesize 0.59 }&{\footnotesize 0.81 }&{\footnotesize \textbf{ 0.97 }}&{\footnotesize \textbf{ 0.01 }}&{\footnotesize \textbf{ 0.02 }}&{\footnotesize 0.41 }&{\footnotesize 0.21 }\\
\hline \end{tabular}\end{table}
  \end{landscape}

\end{document}